\documentclass[hyper,12pt,letterpaper]{JHEP3}

\usepackage{latexsym,amsmath,amsfonts,amssymb}
\usepackage{mathrsfs}
\usepackage[dvips]{graphicx}
\usepackage{bbm}
\usepackage{bm}
\usepackage{subfigure}
\usepackage{paralist}
\newcommand{\slashed}{{\bf\not}}

\numberwithin{equation}{section}

\newcommand{\nn}{\nonumber}
\newcommand{\mat}[1]{\begin{pmatrix} #1 \end{pmatrix}}
\newcommand{\bmat}[1]{\begin{bmatrix} #1 \end{bmatrix}}
\newcommand{\be}{\begin{equation}} 
\newcommand{\ee}{\end{equation}}
\newcommand{\bea}{\begin{equation} \begin{aligned}} \newcommand{\eea}{\end{aligned} \end{equation}}

\newcommand{\bit}{\begin{itemize}} 
\newcommand{\eit}{\end{itemize}}

\newcommand{\cA}{\mathcal{A}}

\newcommand{\cH}{\mathcal{H}}
\newcommand{\cI}{\mathcal{I}}

\newcommand{\cK}{\mathcal{K}}
\newcommand{\CL}{\mathcal{L}}
\newcommand{\cM}{\mathcal{M}}
\newcommand{\cN}{\mathcal{N}}

\newcommand{\cP}{\mathcal{P}}

\newcommand{\cR}{\mathcal{R}}
\newcommand{\cS}{\mathcal{S}}

\newcommand{\Z}{\mathbb{Z}}
\newcommand{\C}{\mathbb{C}}

\renewcommand{\t}{\widetilde }
\renewcommand{\d}{\partial }
\renewcommand{\b}{\bar }

\newcommand{\alphadot}{{\dot\alpha}}

\newcommand{\CLK}{\hat\CL_K}
\newcommand{\CLKt}{\hat\CL_{\b K}}
\newcommand{\CLY}{\hat\CL_Y}
\newcommand{\CLYt}{\hat\CL_{\b Y}}

\newcommand{\half}{{1\over 2}}

\newcommand{\bz}{{\b z}}
\newcommand{\bw}{{\b w}}

\DeclareMathOperator{\Tr}{Tr}

\newcommand{\rs}{{\bf r}}

\newcommand{\jq}{{ q_f}}
\newcommand{\JQ}{{Q_f}}

\title{
The $\cN=1$ Chiral Multiplet  on $T^2 \times S^2$  and Supersymmetric Localization
}

\author{ Cyril~Closset$^{\flat}$ and Itamar~Shamir$^\sharp$\\

{}$^{\flat}$ Simons Center for Geometry and Physics\\ 
Stony Brook University, Stony Brook, NY 11794, USA\\

{}$^{\sharp}$ Department of Particle Physics and Astrophysics \\
Weizmann Institute of Science, Rehovot 76100, Israel
}

\preprint{WIS/12/13-Oct-DPPA}

\abstract{We compute the supersymmetric partition function of an $\mathcal{N}=1$ chiral multiplet coupled to  an external Abelian gauge field on complex manifolds with $T^2 \times S^2$ topology.  The result is locally holomorphic  in the complex structure moduli of $T^2\times S^2$. This computation illustrates in a simple example some recently obtained constraints on the parameter dependence of supersymmetric partition functions.

We also devise a simple method to compute the chiral multiplet partition function on any four-manifold $\cM_4$ preserving two supercharges of opposite chiralities, via supersymmetric localization. In the case of $\cM_4=S^3\times S^1$, we provide a path integral derivation of the previously known result, the elliptic gamma function, which emphasizes its  dependence on the $S^3 \times S^1$ complex structure moduli.
}

\begin{document}

\tableofcontents


\section{Introduction}

$\cN=1$ supersymmetric field theories with an $R$-symmetry can be defined on a compact manifold $\cM_4$ while preserving at least one supercharge, if and only if $\cM_4$ is  Hermitian \cite{Dumitrescu:2012ha,Klare:2012gn}.   It was shown recently \cite{Closset:2013vra} that the partition function $Z_{\cM_4}$ of such theories is independent of the Hermitian metric on $\mathcal{M}_4$ and that it depends holomorphically on the complex structure parameters of the underlying complex manifold.

We initiate the study of supersymmetric partition functions on $T^2 \times S^2$ by computing the partition function  $Z_{T^2 \times S^2}^\Phi$ of an $\cN=1$ chiral multiplet $\Phi$ coupled to an external vector multiplet. (The inclusion of dynamical gauge fields will be discussed elsewhere.) Our result provides a new concrete example to illustrate  the general properties of supersymmetric partition functions obtained in \cite{Closset:2013vra}. In particular, the partition function $Z_{T^2 \times S^2}^\Phi$  is locally holomorphic in the complex structure moduli of $T^2 \times S^2$.

We will work with a two-dimensional moduli space of complex structures on $T^2 \times S^2$.  Consider $\C \times S^2$ with coordinates~\footnote{The coordinate $z$ covers the sphere except for the south pole, where we change coordinates to $z'= {1/z}$.} $w, z$, with the identifications
\be\label{cs S2T2}
(w, z) \sim (w+2\pi, e^{2\pi i\alpha} z) \sim (w+2\pi \tau , e^{2\pi i \beta} z)~.
\ee
Here $\tau= \tau_1 + i \tau_2$ with $\tau_2>0$  is the standard modular parameter of $T^2$, while $\alpha$, $\beta$ are two real parameters which rotate the $S^2$ as it goes around the periods of the torus. The quotient space is diffeomorphic to $T^2\times S^2$. The two complex parameters
\be\label{cs params T2S2}
 \tau= \tau_1 + i \tau_2~, \qquad\quad \sigma =  \tau\alpha -\beta~,
\ee
are the complex structure moduli. There exist additional families of complex structures on $T^2 \times S^2$   \cite{Suwa:1969}. The family that we consider here is complete, in the sense that for generic values of the moduli \eqref{cs params T2S2} the allowed deformations still lie in the same family.

On $T^2 \times S^2$ with the above complex structure, it is possible to preserve (at most) two supercharges of opposite $R$-charge for any value of \eqref{cs params T2S2}.~\footnote{For theories with a Ferrara-Zumino (FZ) supercurrent multiplet, this point was mentioned in \cite{Samtleben:2012gy}.} An important feature of the corresponding supergravity backgrounds \cite{Dumitrescu:2012ha, Closset:2013vra, Festuccia:2011ws} is that  the $R$-symmetry background gauge field has one unit of magnetic flux through the $S^2$. A consequence of this $R$-symmetry monopole is that the supercharges commute with the isometries of the sphere. Another consequence is that we are only allowed to consider fields of integer $R$-charges.~\footnote{Yet another consequence is that the dimensional reduction of  the $T^2 \times S^2$ background to $S^1 \times S^2$ does not correspond to the  3d $\cN=2$ ``superconformal index'' background of \cite{Imamura:2011su}, which can preserve four supercharges, but rather to a different $S^1\times S^2$ background with one unit of magnetic flux. See \cite{Closset:2013vra, Hristov:2013spa} for related discussions of $S^1 \times S^2$ backgrounds.}

Consider an $\cN=1$ chiral multiplet $\Phi$ of $R$-charge $r$ and $\JQ$-charge $\jq$, where $\JQ$ is an Abelian flavor symmetry $U(1)_f$. 
We couple $\Phi$ to a background $U(1)_f$ real vector multiplet in the most general way that preserves the two supercharges \cite{Dumitrescu:2012ha, Closset:2013vra}. The $U(1)_f$ background gauge field has flux $g\in \Z$ through the $S^2$ and flat connections $a_x$, $a_y$ along the one-cycles of the torus. Let us define the shifted $R$-charge $\rs=  r+ \jq \, g$
and the fugacities
\begin{align}  \label{def fugacitis}
q &= e^{2\pi i \tau}\, , & x &= e^{2\pi i(\tau \alpha - \beta)}\,, & t &= e^{2\pi i (\tau a_x - a_y)}~.
\end{align}
The shifted $R$-charge $\rs$ must also be integer. For $\rs>1$,  we find
\begin{align}  \label{result}
Z^{\Phi}_{\rs>1}(q,x,t) &= \left( \frac{q^{1\over 12}}{\sqrt{t^\jq}} \right)^{\rs -1 }\prod_{m= -{\rs\over 2}+1}^{{\rs\over 2}-1} \prod_{k \geq 0} \left(1-  q^{k+1} x^{ -m } t^{-q_f} \right)  \left(1-  q^k x^{ m } t^{q_f} \right)~.
\end{align}%
 Similarly, for $\rs \leq 1$~:
\bea\label{result ii}
&Z^\Phi_{\rs=1}(q,x,t)=1~, \cr
&Z^{\Phi}_{\rs<1}(q,x,t) = \left( \frac{\sqrt{t^\jq}}{q^{1\over12}} \right)^{|\rs| + 1 }\prod_{m= -{|\rs|\over 2}}^{{|\rs|\over 2}} \prod_{k \geq 0} \frac{1}{\left(1-  q^{k+1} x^{ -m } t^{-q_f} \right)\left(1-  q^k x^{ m } t^{q_f} \right)}~.
\eea

We will present two independent derivations of (\ref{result}), (\ref{result ii}). The first method is to compute the partition function by canonical quantization. We define a supersymmetric index for states on $S^1\times S^2$,
\be\label{def S2T2 index} 
\cI_{T^2 \times S^2}= \mathrm{Tr}\left( (-1)^F e^{-2\pi H}\right)~.
\ee
Here $F$ is the fermion number and $H$ is the Hamiltonian with respect to the ``time'' direction along the remaining $S^1$. The Hamiltonian in \eqref{def S2T2 index}  is computed at a generic point on the moduli space of complex structures. On states that contribute to the index, it takes the form
\begin{align}  \label{H short mult}
H = -i \left(\tau P + \sigma J_3 + (\tau a_x - a_y) \JQ\right)~,
\end{align}
with $P$ the momentum along the spatial $S^1$ and $J_3$  the Cartan of the $SO(3)$ isometry of the $S^2$. Such states are thus weighted by $q^P x^{J_3} t^{\JQ}$ in \eqref{def S2T2 index}. 
One could also take the more traditional point of view of computing the index with the ``round'' metric, corresponding to the special point $\tau=i, \sigma=0$ in the complex structure moduli space, while $q, x, t$ would be introduced as fugacities for operators commuting with the supercharges. The advantage of our approach is that the geometrical meaning of the fugacities $q, x, t$ is clear from the onset. 

The index (\ref{def S2T2 index}) coincides with the supersymmetric partition function on $T^2\times S^2$ up to local terms. 
To compute it,  we dimensionally reduce the theory on the sphere. This leads to an infinite tower of decoupled $\cN=(0,2)$ supersymmetric theories on the torus. Only states in short $\cN=(0,2)$ multiplets contribute to (\ref{def S2T2 index}), and the  effective Hamiltonian \eqref{H short mult} acting on those states encodes the dependence on the complex structure moduli $\tau, \sigma$ of $T^2 \times S^2$. The index (\ref{def S2T2 index})   reduces to a Witten index  on $T^2$, also known as an  $\cN=(0,2)$ elliptic genus \cite{Witten:1982df,Witten:1986bf}. Our results for the index overlap with some recent work on elliptic genera \cite{Benini:2013nda,Benini:2013xpa}.

Our second derivation of  (\ref{result}), (\ref{result ii}) is a path integral computation.  
We take advantage of the two Killing spinors present on our geometric background to rewrite the $\cN=1$ chiral multiplet in terms of more convenient variables. 
We then present a general method, based on supersymmetric localization (see for instance \cite{Pestun:2007rz, Kapustin:2009kz,Hama:2011ea}), to compute the  chiral multiplet one-loop determinant on {\it any} four-manifold $\cM_4$ preserving two supercharges.  In particular, in the case $\cM_4= T^2 \times S^2$ we confirm the result  (\ref{result}), (\ref{result ii}).

We also study the partition function of a chiral multiplet on $S^3\times S^1$ using that same localization method. This provides a new derivation of a well-known result. $Z^\Phi_{S^3\times S^1}$ is given by an elliptic gamma function  \cite{Romelsberger:2007ec, Dolan:2008qi}, whose natural domain of definition is the complex structure moduli space of $S^3\times S^1$ \cite{Closset:2013vra}.~\footnote{The computation of  $Z^\Phi_{S^3\times S^1}$ was also reconsidered recently in  \cite{Gerchkovitz:2013zra} from the canonical quantization point of view.}

This paper is organized as follows. In  section \ref{sec: susy with two supercharges}, we review some necessary background about supersymmetry on curved spaces, while we study the $T^2 \times S^2$ background more thoroughly  in section \ref{complex_st}. We derive the result \eqref{result}, \eqref{result ii} in section \ref{sec: compute index} by computing the supersymmetric index \eqref{def S2T2 index} for an $\cN=1$ chiral multiplet, and we briefly describe some of its interesting properties. In section \ref{sec: chiral mult Z}, we propose a general method to compute the chiral multiplet partition function on any background with two supercharges, and we apply this method to the cases of $T^2\times S^2$ and $S^3\times S^1$. Some useful additional material is presented in Appendices \ref{App: monopole harmonics}, \ref{App: Pairing} and \ref{App: S3S1}.


\section{Curved Space Supersymmetry and the $\cN=1$ Chiral Multiplet}\label{sec: susy with two supercharges}
In this section, we review curved space rigid supersymmetry for $\cN=1$ supersymmetric theories with an $R$-symmetry, focussing  on the case of two supercharges of opposite chiralities. We discuss in detail the $\cN=1$ chiral multiplet coupled to an external gauge multiplet. 

\subsection{Background Supergravity Multiplet}
Curved space supersymmetry on compact manifolds is best understood as a rigid limit of off-shell supergravity \cite{Festuccia:2011ws}. Four-dimensional $\cN=1$ supersymmetric theories with an exact $R$-symmetry possess a supercurrent multiplet, called the $\cR$-multipet, whose bottom component is the conserved $R$-symmetry current  \cite{Sohnius:1981tp,Komargodski:2010rb, Dumitrescu:2011iu}. The $\cR$-multiplet  couples to the new minimal supergravity multiplet of \cite{Sohnius:1981tp}, which contains the metric $g_{\mu\nu}$, a gravitino, and two auxiliary fields $A_\mu$ and $V_\mu$. The field $A_\mu$ is an $R$-symmetry gauge field while $V_\mu$ is a vector which satisfies $\nabla_\mu V^\mu=0$. Rigid supersymmetry on a compact four-manifold $\cM_4$ corresponds to a choice of supersymmetric background values for $g_{\mu\nu}, A_\mu, V_\mu$ \cite{Festuccia:2011ws}, such that the (generalized) Killing spinor equations 
\bea\label{KSE}
&(\nabla_\mu  - i A_\mu )\zeta = - i V_\mu \zeta - i V^\nu \sigma_{\mu\nu} \zeta~,\cr
& (\nabla_\mu  + i A_\mu ) \t \zeta = i V_\mu \t \zeta + i V^\nu \t \sigma_{\mu\nu} \t \zeta~,
\eea
admit at least one non-trivial solution, which we can take to be $\zeta$. The Killing spinors $\zeta_\alpha$ and $\t\zeta^\alphadot$ are spinors of $R$-charge $+ 1$ and $-1$, respectively, of opposite chiralities. Any non-trivial solution is nowhere vanishing.

In \cite{Dumitrescu:2012ha, Klare:2012gn}, it was shown that the existence of a single Killing spinor $\zeta$ is equivalent to the existence of a complex structure on $\cM_4$ compatible with the metric. Given $\zeta$, the metric-compatible complex structure is given explicitly by~\footnote{We are following the conventions of  \cite{Closset:2013vra}, which differ from the ones of \cite{Dumitrescu:2012ha} by some signs. However, our $R$-symmetry gauge field $A_\mu$ is defined as in \cite{Dumitrescu:2012ha}, with $A_\mu^{(R)}= A_\mu -{3\over 2} V_\mu$ the gauge field used in \cite{Closset:2013vra}.}
\be\label{csfromzeta}
{J^\mu}_\nu = -{2 i \over |\zeta|^2} \zeta^\dagger {\sigma^\mu}_\nu \zeta ~.
\ee
Conversely, given a complex manifold with a choice of a Hermitian metric, one can solve for the auxiliary fields $A_\mu$, $V_\mu$ and for $\zeta$ explicitly \cite{Dumitrescu:2012ha}.
In this paper we will focus on backgrounds preserving at least {\it two supercharges of opposite chiralities}, $\delta_\zeta$ and $\delta_{\t\zeta}$.
Given two Killing spinors  $\zeta$ and $\t\zeta$, there exists a complex Killing vector \cite{Dumitrescu:2012ha}
\be\label{Kvec}
K= \zeta \sigma^\mu \t\zeta \, \d_\mu\, .  
\ee
This vector  is anti-holomorphic with respect to (\ref{csfromzeta}).
Moreover, we will assume that $K$  commutes with its complex conjugate, $[K,  K^\dagger]=0$.~\footnote{This is what we mean by ``two supercharges'' throughout this paper. The case $[K, K^\dagger]\neq 0$ was analysed in  \cite{Dumitrescu:2012ha}.} Such backgrounds are torus fibrations over a Riemann surface $\Sigma$. We can pick complex coordinates $w, z$  adapted to the complex structure (\ref{csfromzeta}) such that $K= \d_{\b w}$. The Hermitian metric  is locally given by
\be\label{2Q backgd i}
ds^2 = \Omega^2(z, \b z) \Big((dw + h (z,\b z)  d z)(d\b w + \b h (z,\b z)  d \b z) + c^2(z,\b z)  dz d\b z \Big)\, .
\ee
Note that, in general, the real and imaginary parts of $K$ do not have closed orbits, but   they are  instead part of a larger $U(1)^3$ isometry. $\cM_4$ is therefore a $T^2$ fibration (by choosing a $T^2$ out of the $U(1)^3$ isometry orbits), but not necessarily a holomorphic one. (The fibration is holomorphic only when the orbits of $K$ are bona fide tori.)~\footnote{See section 5.1 of \cite{Closset:2012ru} for a similar discussion in three dimensions.}
The background supergravity fields $A_\mu, V_\mu$ read
\bea\label{2Q backgd ii}
& V_\mu  = \half \nabla_\nu {J^\nu}_\mu + \kappa K_\mu \, , \cr
&A_\mu =  - {1\over 4} {J_\mu}^\nu\d_\nu \log \sqrt{g} +{1\over 4} ({\delta_\mu}^\nu+ i {J_\mu}^\nu) \nabla_\sigma {J^\sigma}_\nu +{3\over 2} \kappa K_\mu -{i\over 2} \d_\mu \log s ~,
\eea
The expression for $A_\mu$ is only valid in complex coordinates ---in particular, $\sqrt{g}= {1\over 4}\Omega^4 \, c^2$ is the determinant of the Hermitian metric (\ref{2Q backgd i}). The function $\kappa$ is such that $K^\mu \d_\mu\kappa=0$, but otherwise arbitrary. In the holomorphic frame 
\be\label{standard frame}
e^1=\Omega(z,\b z)(dw + h(z,\b z)  d z) \, ,\qquad  e^2 = \Omega(z,\b z) c(z,\b z) dz\, ,
\ee
 the Killing spinors read
\be\label{Killingspinors}
\zeta_\alpha = \sqrt{\frac{s}{2}} \left(  \begin{matrix}1 \cr   0 \end{matrix}\right)\, , \quad\qquad  
\t \zeta^\alphadot =   {\Omega \over \sqrt{2s}}  \left( \begin{matrix} -1 \cr   0 \end{matrix}\right)\,  
\ee
From the Killing spinor $\zeta_\alpha$, we can also construct a holomorphic two-form
\be\label{P holo}
P =\zeta \sigma_{\mu\nu}\zeta \, dx^\mu \wedge dx^\nu =s\, g^{1\over 4} \, dw \wedge dz~.
\ee
Note that $s$ is a nowhere-vanishing global section  of $\cK \otimes L^2$, with $\cK$ the canonical line bundle and $L$ the $R$-symmetry line bundle. Therefore, the line bundle $\cK \otimes L^2$ is trivial and we can identify $L \cong \cK^{-\half}$ up to a trivial line bundle  \cite{Dumitrescu:2012ha}. 

\subsection{Background Vector Multiplet}\label{subsec: backgd amu}
In addition to the supergravity background (\ref{2Q backgd i}), (\ref{2Q backgd ii}), it is  natural to consider background gauge fields coupling to conserved currents (whenever the field theory has any global symmetry). 
For simplicity, consider a background vector multiplet  for an Abelian  symmetry $U(1)_f$. It has bosonic components $v_\mu$ and $D$, with $v_\mu$ the $U(1)_f$  gauge field. Let us define its field strength
\be\label{def fmn for v}
f_{\mu\nu}= \d_\mu v_\nu - \d_\nu v_\mu \, .
\ee
In order to preserve the same Killing spinors (\ref{Killingspinors}) as the supergravity background (\ref{2Q backgd i}), (\ref{2Q backgd ii}), the background fields $v_\mu$, $D$ have to satisfy   \cite{Closset:2013vra} 
\be\label{background V}
 f_{\b w\b z}= f_{w\b w}=f_{z\b w} =0\, , \qquad D=-{2 i \over \Omega^2 c^2}  (f_{z\b z} - h f_{w \b z})~,
\ee
using our adapted coordinates $w, z$ and the metric (\ref{2Q backgd i}).  For $v_\mu$ real,  $f_{z\b z}$ is the only component of the field strength that can be turned on while preserving two supercharges.  

\subsection{Supersymmetry Transformations and Supersymmetric Lagrangian}\label{subsec: chiral mult}
Let $\delta_\zeta$ and $\delta_{\t\zeta}$ be the two supersymmetries associated to the Killing spinors $\zeta$ and $\t\zeta$.  They satisfy the supersymmetry algebra \cite{Dumitrescu:2012ha}
\be\label{susy alg}
\delta_\zeta^2=0\, , \qquad \delta_{\t\zeta}^2=0\, , \qquad \{\delta_{\zeta},\delta_{\t\zeta}\}= 2 i \hat\CL_K\, .
\ee
Here $K$ is the Killing vector (\ref{Kvec}).
The twisted Lie derivative $\hat\CL_K$ along $K$ acts as
\be\label{defLprime}
\hat\CL_K   = \CL_K  -i K^\mu (  r A_\mu +  \jq v_\mu) 
\ee
on any field of $R$-charge $r$ and $\JQ$-charge $\jq$, with $\CL_K$  the ordinary Lie derivative.~\footnote{Note that the background gauge field $v_\mu$ appears in the supersymmetry algebra (\ref{susy alg}). Technically, this is because we work with the Wess-Zumino gauge. }

We consider a chiral multiplet $\Phi=(\phi, \psi, F)$ of $R$-charge $r$  and $\JQ$-charge $\jq$, in the supersymmetric background   (\ref{2Q backgd i}), (\ref{2Q backgd ii}), (\ref{background V}). Let us define the covariant derivative
\be \label{covder}
D_\mu = \nabla_\mu - i r A_\mu - i \jq v_\mu~,
\ee
acting on fields of charges $r$ and $\jq$.
The curved space supersymmetry transformations read \cite{Festuccia:2011ws}
\bea\label{susychiralfdgauged}
&\delta \phi = \sqrt2 \zeta \psi~,\cr
&\delta \psi_\alpha = \sqrt 2 \zeta_\alpha F + \sqrt 2 i (\sigma^\mu \t \zeta)_\alpha \, D_\mu \phi~, \cr
&\delta F =\sqrt 2 i  D_{\mu}(\t \zeta \t \sigma^\mu \psi).
\eea
Similarly, for an anti-chiral multiplet $\t\Phi=(\t\phi, \t\psi,\t F)$ of $R$-charge $-r$ and $\JQ$-charge  $-\jq$, we have
\bea\label{susyantichiralfdgauged}
&\delta \t\phi = \sqrt2 \t\zeta \t\psi~,\cr
&\delta \t\psi^{\dot\alpha} = \sqrt 2 \t\zeta^{\dot\alpha} \t F + \sqrt 2 i (\t \sigma^\mu  \zeta)^{\dot\alpha} \, D_\mu \t\phi~, \cr
&\delta \t F =\sqrt 2 i D_{\mu}( \zeta \sigma^\mu \t \psi).
\eea
One can check that these supersymmetry transformations realize the supersymmetry algebra  (\ref{susy alg}) if and only if $\zeta$, $\t\zeta$ are solutions of (\ref{KSE}).
One can construct the curved-space generalization of the canonical kinetic term for a chiral multiplet coupled to a background gauge field.
It is given by the following supersymmetric Lagrangian~:
\begin{align}\label{Lag Phi}
\mathscr{L}_{\Phi\t\Phi} &=  \;   D_\mu \t \phi D^\mu \phi   + i \t \psi \t\sigma^\mu D_\mu \psi - F\t F 
+V^{\mu}\big(iD_{\mu} \tilde{\phi}\, \phi  -i \tilde{\phi} D_{\mu} \phi +\half \tilde{\psi}\tilde{\sigma}_{\mu} \psi\big)  \nonumber \\
& \quad   -{r\over 4} (R-6V^{\mu}V_{\mu})\, \t \phi \phi  +   \jq  D \t\phi  \phi \, .
\end{align}
Here $R$ is the Ricci scalar on $\cM_4$,~\footnote{We follow the Riemannian geometry conventions of \cite{Dumitrescu:2012ha}. In particular, $R < 0$ for a round sphere.} and the various background fields take their supersymmetric values.

\subsection{A Comment on Global Anomalies}
The theory of a single chiral multiplet that we are considering  has cubic and mixed $U(1)$-gravitional anomalies for its $U(1)_f$ and $U(1)_R$ symmetries~:
\be
\Tr( R^3)= (r-1)^3~, \quad \Tr (Q_f^3)= q_f^3~, \quad 
\Tr (R)= r-1~, \quad \Tr (\JQ)= \jq~, 
\ee
and similarly for $\Tr (R^2 Q_f)$ and $\Tr (Q_f^2 R)$. This could result in a violation of current conservation in a non-trivial background. Let us define the topological densities
\be
\cP^{(f)}= \epsilon^{\mu\nu\rho\sigma} f_{\mu\nu} f_{\rho\sigma}~, \qquad
\cP^{(R)}=  \epsilon^{\mu\nu\rho\sigma} F^{(R)}_{\mu\nu} F^{(R)}_{\rho\sigma}~, \qquad
\cP^{(g)}=  \epsilon^{\mu\nu\rho\sigma} R_{\mu\nu\kappa\lambda}{R_{\rho\sigma}}^{\kappa\lambda}~.
\ee
Here $f_{\mu\nu}$ is the $U(1)_f$ field strength, $F_{\mu\nu}^{(R)}$ is the field strength of $A_\mu^{(R)}= A_\mu -{3\over 2}V_\mu$, which couples to the $R$-symmetry current, and $R_{\mu\nu\rho\sigma}$ is the Riemann tensor.

 In the presence of two supercharges, we have
\be
\cP^{(f)}= 0~, \qquad \quad \cP^{(R)} = {3\over 8} \cP^{(g)}~.
\ee
The first equality  follows directly from \eqref{background V} and the second relation was derived in \cite{Cassani:2013dba} (it is also easily checked from \eqref{2Q backgd i} and \eqref{2Q backgd ii}). Since $\cP^{(f)}=0$, the $\Tr(Q_f^3)$ and $\Tr(Q_f^2 R)$ cubic anomalies do not lead to any violation of current conservation. On the other hand, $\cP^{(g)}$  will be non-zero in general. However, one can always write $\cP^{(g)}= \nabla_\mu X^\mu$ for $X^\mu$ a non-covariant quantity, so that the properly shifted currents are conserved. In general, we can thus preserve $U(1)_f$ and $U(1)_R$ at the expense of general covariance \cite{AlvarezGaume:1983ig}. (It was also noted in \cite{Cassani:2013dba} that the integral of $\cP^{(g)}$ vanishes on backgrounds with two supercharges, so that all the integrated anomalies vanish.)

In any case, for all the backgrounds considered in this paper, one can actually show that $\cP^{(g)}=0$, and therefore the $U(1)_f$ and $U(1)_R$ currents are properly conserved.~\footnote{One can see that $\cP^{(g)}=0$ if $h=\b h =0$, with $h$ the metric function appearing in \eqref{2Q backgd i}. One can also check that  $\cP^{(g)}=0$ for the $S^3 \times S^1$ background that we will consider in section \ref{subsec: S3 S1 Z}.} Therefore, we will not need to worry about these global anomalies in the following.

\section{Complex Structures and Supersymmetry on $T^2 \times S^2$} \label{complex_st}
Any complex four-manifold with $T^2 \times S^2$ topology is a ruled surface of genus one \cite{Suwa:1969}.  Such surfaces have been classified \cite{ Suwa:1969,Atiyah:1955}.~\footnote{A ruled surfaces of genus one is a $\C P^1$ fiber bundle over a non-singular elliptic curve $\Sigma_1$. The classification of such surfaces follows from the classification of one-dimensional affine fiber bundles over $\Sigma_1$, with the ruled surfaces obtained by projectivisation \cite{Atiyah:1955}. It was shown in \cite{Suwa:1969} that there are three classes of complex structures on $T^2 \times S^2$, denoted by $\cS$, $S_{2n}$ with $n$ any positive integer, and $A_0$. In this paper we consider the class $\cS$, which corresponds to ruled surfaces obtained from degree zero holomorphic line bundles over $\Sigma_1$. 
}
In this paper, we consider a two-parameter family of complex manifolds obtained as quotients of $\C\times S^2$. Let $w, z$ be the complex coordinates on $\C$ and $S^2$, respectively. We consider the identifications
\be\label{complexstructTtwo}
(w, z) \sim (w+2\pi, e^{2\pi i\alpha} z)~, \qquad (w, z) \sim (w+ 2\pi \tau , e^{2\pi i \beta} z)~.
\ee
The quotient space is diffeomorphic to $T^2 \times S^2$.
Here  $\tau=\tau_1 + i \tau_2$ is the modular parameter of the torus, and $\alpha$, $\beta$ are real parameters subject to the identifications $\alpha\sim \alpha + 1$, $\beta\sim \beta +1$. We also introduce the complex parameter $\sigma = \tau \alpha - \beta$.
Two choices of complex structure moduli $\tau, \sigma$ and $\tau', \sigma'$ are equivalent if they give rise to the same identifications (\ref{complexstructTtwo}). These equivalences are generated by~:
\bea\label{Gcsi}
& S \; :  \; \; (\tau, \sigma) \mapsto \left(-{1\over \tau},\, {\sigma \over \tau} \right)\, , \qquad\quad
& T \; :  \; \; (\tau, \sigma) \mapsto (\tau+1, \, \sigma)\, , \cr
& U \; :  \; \; (\tau, \sigma) \mapsto (\tau, \, \sigma +\tau )\, , \qquad\quad
& V \; : \; \; (\tau, \sigma) \mapsto (\tau, \,  \sigma + 1)\, ,
\eea
These transformations generate a subgroup of $PSL(3,\Z)$, whose corresponding $SL(3,\Z)$ matrices are 
\be\label{subgroupSLthree}
S=     \mat{0 &1& 0\cr-1 &0&0\cr 0& 0 & 1 }~,  \qquad
 T=     \mat{1 &1& 0\cr0 &1&0\cr 0& 0 & 1 }~, \qquad U=    \mat{1 &0& 0\cr0 &1&0\cr 1& 0 & 1 }~, \qquad
 V=    \mat{ 1&0& 0\cr0 &1&0\cr 0& 1 & 1 }~.
\ee

It is convenient to introduce real coordinates $x$, $y$, $\theta$, $\varphi$ on $T^2 \times S^2$, where $x$, $y$ are torus coordinates of period $2\pi$, and $\theta \in [0, \pi]$, $\varphi\in [0, 2\pi)$ are the standard angles on the sphere. 
The  complex structure on the quotient (\ref{complexstructTtwo}) can be realized by the complex coordinates
\be\label{cc on S2T2}
w = x + \tau y \, , \qquad z= \tan{\theta\over 2} e^{i (\varphi +  \alpha x + \beta y)}~,
\ee
where the identifications (\ref{complexstructTtwo}) correspond to $(x,y)\sim (x+ 2\pi, y)$ and $(x,y)\sim (x, y+2\pi)$ on the real torus. 
The generators (\ref{Gcsi}) of non-trivial identifications on the complex structure moduli space correspond to large diffeomorphisms of the underlying real manifold, which are given by the matrices (\ref{subgroupSLthree}) acting on the coordinates $(x, y, \varphi)$ in the obvious way.

\subsection{Supergravity Background with Round Metric}\label{subsec: S2T2 back round}
One can preserve two supercharges on $T^2\times S^2$  for any choice of complex structure (\ref{complexstructTtwo}). 
We consider the metric
\be\label{metricprod}
ds^2 = d w d\b w + {4 \over (1 + z \b z)^2} dz d \b z\, ,
\ee
which is compatible with the identification (\ref{complexstructTtwo}). Note that for generic values of the complex structure moduli, this metric has three real Killing vectors, corresponding to $\d_w$, $\d_\bw$ and $z\d_z -\bz \d_z$. (The additional Killing vectors $K_\pm$ in \eqref{3 KV} are not globally defined unless $\sigma=0$.)
In particular, we have the anti-holomorphic Killing vector $K= \d_{\b w}$ and we can apply the general formulas (\ref{2Q backgd i}), (\ref{2Q backgd ii}) for the supergravity background fields. 
In terms of the real coordinates (\ref{cc on S2T2}), the metric (\ref{metricprod}) reads
\be
ds^2 = (dx + \tau_1 d y)^2 + \tau_2^2 dy^2  + d\theta^2 + \sin^2\theta (d\varphi + \alpha dx + \beta dy)^2~,
\ee
while the anti-holomorphic Killing vector $K$ is given by
\be\label{K real coord}
K= \d_\bw = {1\over 2 i \tau_2 }\left( \tau\, \d_x -\d_y -\sigma \, \d_\varphi \right)~.
\ee
Formula (\ref{2Q backgd ii}) simplifies greatly because the metric (\ref{metricprod}) is K\"ahler.  We have~\footnote{In complex coordinates, $K_{\mu}dx^{\mu} = \frac{1}{2} d w$ and $A^{(R)} = -\frac{i}{2} \frac{\bar z dz - z d \bar z}{(1+z\bar z)} - \frac{i}{2} d \log s $.}
\begin{align}\label{background S2T2}
V_\mu dx^\mu &= \kappa K_\mu dx^\mu\, , \nonumber \\
A_\mu^{(R)} dx^\mu &= \half (1-\cos\theta) (d\varphi +\alpha dx +\beta dy)   -{i\over 2} \d_\mu \log s\, dx^\mu~,  
\end{align}
with $A_\mu^{(R)}= A_\mu -{3\over 2}V_\mu$.
The salient feature of this background is that it involves a non-trivial $R$-symmetry line bundle $L$, with first Chern class
\be\label{flux dA}
c_1= {1\over 2\pi}\int_{S^2} d A^{(R)} = 1~.
\ee
Any field $\Phi$ of nonzero $R$-charge $r$ must be a well-defined section of the line bundle $L^r$, with transition functions 
\be
\Phi^{(N)} = e^{i r ( \varphi + \alpha x + \beta y)} \Phi^{(S)}
\ee
between the northern $(N)$ and sourthern $(S)$ patches.~\footnote{By default, all quantities are written on the northern patch, $\Phi= \Phi^{(N)}$, with complex coordinates $w,z$. The southern patch has coordinate $w, z'$, with $z'={1\over z}$ on the overlap.}
It follows that the $R$-charge must be {\it integer}, $r\in \Z$.
In the canonical frame (\ref{standard frame}), $e^1 = dw$, $e^2={2\over 1+ |z|^2} dz$,
the Killing spinors are given by (\ref{Killingspinors}).  In order for the holomorphic two-form (\ref{P holo}) to be well-defined,  $s$ must satisfy
\be\label{s transfo T2}
s \sim e^{-2 \pi i \alpha}  s~, \qquad s \sim e^{-2 \pi i \beta}  s
\ee
 under the identifications (\ref{complexstructTtwo}).
We should offset (\ref{s transfo T2}) by some $R$-symmetry transformations, so that $s$ is invariant as we go around the one-cycles of the torus. Therefore, any field $\Phi$ of $R$-charge $r$ has twisted periodicities
\be\label{twisted bd i}
\Phi \sim e^{i \pi r \alpha } \Phi~, \qquad \Phi \sim e^{i \pi r \beta} \Phi~.
\ee
under (\ref{complexstructTtwo}). 
Finally, note that $s$ is a non-trivial section of the canonical line bundle $\cK$ but a trivial section of the total bundle $\cK\otimes L^2$, which is therefore trivial \cite{Dumitrescu:2012ha}. We will take $s=1$ in the following.

Note that the $R$-symmetry flux through the $S^2$ is necessary to preserve supersymmetry on $T^2 \times S^2$. This is in contrast to lower dimensional cases where supersymmetric backgrounds without flux exist on $S^1 \times S^2$ \cite{Imamura:2011su} and $S^2$ \cite{Benini:2012ui,Doroud:2012xw}. Such backrgounds preserve four supercharges and do not admit an uplift to four dimensions \cite{Festuccia:2011ws}.

\subsection{Supersymmetric Background Gauge Field}\label{subsec: backgd V}
Let us also consider a supersymmetric background gauge multiplet for an Abelian symmetry $U(1)_f$, which satisfy (\ref{background V}). We consider a real gauge field for simplicity. The most general such background gauge field preserving the isometries of \eqref{metricprod} is given by 
\bea \label{background a}
v_\mu dx^\mu &= v_w dw + v_\bw d w    -i g \frac{\bar z dz - z d \bar z}{2(1+z\bar z)} \cr
&= a_x dx + a_y dy +{g\over 2} (1-\cos\theta)(d\varphi +\alpha dx+\beta dy)~, 
\eea
up to gauge transformations, while the auxiliary field $D$ is given in (\ref{background V}).
Note that
\be\label{def v vb}
v_w= -{\b \tau a_x- a_y\over 2 i \tau_2}~, \qquad  v_\bw= {\tau a_x- a_y\over 2 i \tau_2}~.
\ee
The real parameters $a_x$, $a_y$ are flat connections, which must be identified by $a_x\sim a_x +1$, $a_y \sim a_y +1$ due to large gauge transformations. It is also convenient to define 
\be\label{def nu}
\nu = \tau a_x - a_y~,
\ee
a line bundle modulus  \cite{Closset:2013vra} analogous to the complex structure modulus $\sigma$. The parameter $g$ is an integer, giving us the quantized flux of  (\ref{background a}) through the sphere. The discussion of the corresponding $U(1)_f$ line bundle is analogous to the discussion of the $U(1)_R$ bundle above. In particular, a field $\Phi$ of $\JQ$-charge $\jq$ has transition function $\Phi^{(N)} = e^{i  g \jq ( \varphi + \alpha x + \beta y)} \Phi^{(S)}$ between the northern and southern patches.
It will be useful to define the shifted $R$-charge
\be\label{shifted R def}
\rs= r + \jq g~.
\ee
In the presence of the background gauge field \eqref{background a}, the twisted boundary conditions \eqref{twisted bd i} for charged fields generalize to~\footnote{It is clear from the transition functions that there must be some twisted  periodicities depending on $\jq g$. This choice of boundary conditions is symmetric between the northern and southern patches.}
\be\label{twisted bd ii}
\Phi \sim e^{i \pi \rs \alpha } \Phi~, \qquad \Phi \sim e^{i \pi \rs \beta} \Phi~,
\ee
under the identifications (\ref{complexstructTtwo}).

\subsection{More General Supersymmetric Background}
We can consider supersymmetric backgrounds on $T^2\times S^2$ with more general metrics than (\ref{metricprod}). Indeed, any Hermitian metric of the local form (\ref{2Q backgd i}) is as good as any other. We can still retain the map (\ref{cc on S2T2}) between real and complex coordinates, and thus the explicit form (\ref{K real coord}) for the Killing vector $K$. Let us note that the real and imaginary parts of the Killing vector $K$ do not close in general, and consequently we have three Killing vectors $\d_x$, $\d_y$ and $\d_\varphi$. Thus we can consider a general background (\ref{2Q backgd i}), (\ref{2Q backgd ii}) with the functions $\Omega(z, \bz)$, $h(z, \bz)$ and $c(z, \bz)$ depending on $|z|^2$ only. One can similarly consider more  general background gauge fields than \eqref{background a} with the same flux $g$  through the $S^2$.

For $\sigma=0$, we need only have  two Killing vectors $\d_x$ and $\d_y$. In this special case, the $T^2\times S^2$ index that we will compute below does not keep track of the $J_3$ quantum number from the sphere.

\section{The $T^2\times S^2$ Index and Canonical Quantization}\label{sec: compute index}
In this section, we compute the $T^2 \times S^2$ partition function of a chiral multiplet as a supersymmetric index (\ref{def S2T2 index}). The first step is to  dimensionally reduce the theory over $S^2$.  Due to the presence  of magnetic flux on the sphere, charged fields must be expanded in so-called monopole spherical harmonics \cite{Wu:1976ge}, which  are reviewed in Appendix \ref{App: monopole harmonics}. The second step is to quantize the resulting theory on the torus. The only contribution to the index will come from a finite number of short multiplets of the $\cN=(0,2)$ supersymmetry on the torus, arising from zero modes of the Dirac operator on the $S^2$ with magnetic flux.

\subsection{Sphere Reduction and $\cN=(0,2)$ Multiplets on $T^2$}
Consider the supersymmetric background with round metric (\ref{metricprod}) discussed in the previous section. We consider a free chiral multiplet of $R$-charge $r$ and $\JQ$-charge $\jq$ in that background.
The bosonic part of the supersymmetric Lagrangian (\ref{Lag Phi}) can be written
\be\label{Lag Phi ii bos}
\mathscr{L}_{{\rm bos}}= 2 (D_w + i \gamma) \t\phi D_\bw \phi  + 2 D_\bw \t\phi (D_w- i \gamma)\phi +\t\phi \left(\Delta^{\rs}_{S^2} +{\rs\over 2}  \right)\phi  - \t F F~,
\ee
where we introduced the notation $\gamma= {3\over 4}\kappa\left(r - {2\over 3}\right)$, with $\kappa$ the ambiguity in the background fields (\ref{background S2T2}), and $\rs$ is the shifted $R$-charge \eqref{shifted R def}. Here and in the remainder of this section, the covariant derivatives along $w, \bw$ only contain the $U(1)_f$ flat connection~:
\be
D_w = \d_w - i \jq v_w~, \quad D_\bw = \d_\bw - i \jq v_\bw~.
\ee
The operator $\Delta^{\rs}_{S^2}$ in (\ref{Lag Phi ii bos}) is the scalar Laplacian on the sphere with a monopole, which is given by (\ref{S2 Lap scalar})  Appendix \ref{App: monopole harmonics}. Note that $R$-charge $r$ only enters through $\gamma$ and the shifted R-charge $\rs$. Indeed, the  scalar field $\phi$ couples to both $U(1)_R$ and $U(1)_f$ through the covariant derivative (\ref{covder}), so that it effectively couples with electric charge $\rs$ to a monopole of unit flux.~\footnote{We also have couplings of $\phi$ to $R$ and $D$ in the second line of (\ref{Lag Phi}), which are crucial for supersymmetry and  lead to the shift of the scalar Laplacian by ${\rs\over 2}$ in \eqref{Lag Phi ii bos}.} Let us also remark that the Laplacian (\ref{S2 Lap scalar}) must be here interpreted in terms of the complex coordinate $z$ in (\ref{cc on S2T2}), effectively shifting the angle $\varphi$ to $\varphi +\alpha x + \beta y$ in the definition of the monopole harmonics. (The monopole harmonics are then a complete basis of sections of the monopole line bundle on $T^2 \times S^2$ introduced in section \ref{subsec: S2T2 back round}.) We expand the field $\phi$ in scalar monopole harmonics,
\be\label{expansion phi}
\phi(w,\bar{w},z,\bar{z})= \sum_{j,m}  a_{jm}(w,\bar{w}) Y_{{\rs} \, j m}(z,\bar{z})~, 
\ee
where the sum is over $j= \frac{|\rs|}{2}, \frac{|\rs|}{2}+1,\ldots$, $m=-j,\ldots,j$. Similarly, we expand the auxiliary field $F$ in monopole harmonics of electric charge $\rs-2$~:
\be
F(w,\bar{w},z,\bar{z}) = \sum_{j,m}  f_{jm}(w,\bar{w}) Y_{{\rs-2}\, j m}(z,\bar{z})~, 
\ee
with $j= \frac{|\rs-2|}{2}, \frac{|\rs-2|}{2}+1,\ldots$,  $m=-j,\ldots,j$. We similarly expand the bosonic fields of the antichiral multiplet according to $\t\phi = \sum \t a_{jm} Y_{\rs\, jm}^\dagger$ and $\t F = \sum \t f_{jm} Y_{\rs-2\, jm}^\dagger$. 
Note that there is a mismatch in the $SO(3)$ representations that appear in the expansion of $\phi$ and $F$ for $\rs \neq 1$. When $\rs>1$, all values of $j\geq \frac{\rs}{2}$ exist for both fields but one additional representation with $j=\frac{\rs}{2}-1$ is found for $F$. When $\rs<1$, it is $\phi$ which has one unmatched representation, with $j=-\frac{\rs}{2}$.

The fermionic part of the supersymmetric Lagrangian (\ref{Lag Phi}) is
\be\label{Lag Phi ii fer}
 \mathscr{L}_{{\rm fer}}  = -2 i \tilde{\psi}^{\dot{\beta}} \begin{pmatrix} D_\bw & 0 \\  0 &  -(D_w - i \gamma) \end{pmatrix}\psi_{\alpha} - \tilde{\psi}^{\dot{\beta}} {(-i \slashed{\nabla}^{\rs}_{S^2})_{\dot{\beta}}}^{\alpha} \psi_{\alpha}~.
\ee
with $\gamma$ defined above, while the explicit form of the Dirac operator in a monopole background, $-i \slashed{\nabla}^{\rs}_{S^2}$, is given in (\ref{Dirac op on S2}). 
The eigenvalues $\lambda_{\rs j}$  of $-i \slashed{\nabla}^{\rs}_{S^2}$ are given in  (\ref{eigenvalues Dirac}). The important thing for us is that there are fermionic zero-modes, 
\be
\lambda_{\rs j}=0  \qquad \Leftrightarrow \qquad  j =  \frac{|\rs-1|}{2}-\frac{1}{2}~.
\ee
We expand the fermions in spinors spherical harmonics of electric charge $\rs-1$.
For $\rs >1$, it is convenient to write
\be \label{psi-exp}
\psi_{\alpha} = \sum_{jm}\begin{pmatrix} b_{jm} Y_{\rs-2\, jm} \\ -c_{jm} Y_{\rs\,jm} \end{pmatrix}  +  \sum_m  \begin{pmatrix} b_{j_0 m} Y_{\rs-2\,jm} \\ 0 \end{pmatrix}~,
\ee
The second sum comes from the zero modes, with $j_0=\frac{\rs}{2}-1$.
For $\rs<1$, we similarly write
\be\label{psi-exp ii}
\psi_{\alpha} =  \sum_{jm}\begin{pmatrix} b_{jm} Y_{\rs-2\, jm} \\ -c_{jm} Y_{\rs\, jm} \end{pmatrix} + \sum_m  \begin{pmatrix} 0 \\ -c_{j_0 m} Y_{\rs\, jm} \end{pmatrix}~,
\ee
with $j_0 = {|\rs|\over 2}$. For the fermion $\t\psi$ of opposite chirality, we  take
\be
\tilde{\psi}^{\dot{\alpha}} = -\sum_{jm}\begin{pmatrix} \tilde{b}_{jm} Y^\dagger_{\rs-2\, jm} \\  \tilde{c}_{jm} Y^\dagger_{\rs\, jm}  \end{pmatrix} -\sum_m \begin{pmatrix} \tilde{b}_{j_0 m} Y^\dagger_{\rs-2\, j_0m} \\  0  \end{pmatrix} 
\ee
in the case $\rs >1$, and similarly for $\rs<1$.

Plugging these mode expansions into the Lagrangian (\ref{Lag Phi ii bos}),  (\ref{Lag Phi ii fer}), we obtain a decoupled theory for each angular momentum $j, m$. For simplicity of notation, we will drop the subscripts $j,m$ on the modes $a,b,c,f$ in the following. The modes organize themselves into representations of an $\cN=(0,2)$ supersymmetry algebra on the torus. Let us consider each case separately~:

\subsubsection{Long Multiplets}
Consider first the case when $j, m$ do not correspond to a zero mode, $\lambda \neq 0$. One obtains long multiplets $(a,b,c,f)$ and $(\t a, \t b,\t c,\t f)$, whose supersymmetry variations follow from (\ref{susychiralfdgauged}) and (\ref{susyantichiralfdgauged})~: 
\bea\label{S-var-csf}
\delta a &= c~, & \qquad \t \delta a &= 0~, &\qquad \delta \t a &=0~, &\qquad \t\delta \t a &= \t c~, \\
\delta b &= f~, &\qquad \t \delta b &= i \lambda a~, &\qquad \delta \t b &= -i \lambda \t a~, &\qquad \t \delta \t b &= \t f~,   \\
\delta c &=0~, &\qquad \t \delta c &=  2iD_{\bar{w}} a~, &\qquad \delta \t c &= 2 iD_\bw \t a~, &\qquad \t \delta \t c &=0~,  \\
\delta f &=0~, &\qquad \t \delta f &=   2i D_{\bar{w}} b - i \lambda c ~, &\qquad \delta \t f &= 2i D_\bw \t b + i \lambda \t c~, &\qquad \t \delta \t f &=0~.
\eea
Here we use the notation $\delta = \delta_{\zeta}$ and $\t \delta = \delta_{\t \zeta}$.
The transformations (\ref{S-var-csf})  realize a two-dimensional $\cN=(0,2)$ algebra, 
\be
\delta^2=0~, \qquad \t\delta^2=0~, \qquad \{ \delta , \t \delta\} = 2 i D_{\bar{w}}~.
\ee
The Lagrangian for the $(a,b,c,f)$ multiplet is given by~\footnote{Note the dependence on the ambiguity parameter $\gamma$, which is now an arbitrary constant (for simplicity we assumed a constant $\kappa$, thus preserving all the isometries of the background). In the following we tacitly assume that $\gamma$ is real, although the generalization to complex $\gamma$ is straightforward.}
\bea\label{L abcf}
\mathscr{L}_{(a,b,c,f)} &= 4 D_{\bar{w}} \tilde{a} D_w  a + \lambda^2 \tilde{a} a  + 2 i\gamma (\tilde{a} D_{\bar{w}} a - a D_{\bar{w}} \tilde{a}) - \tilde{f}f  \\ 
& \quad + 2i \tilde{b} D_{\bar{w}} b + 2i \tilde{c} D_{w} c +  2\gamma \tilde{c}c - i \lambda (\tilde{b}c + \tilde{c}b)~.
\eea
One  easily checks that (\ref{L abcf}) is supersymmetric under (\ref{S-var-csf}).

\subsubsection{Short Multiplets}
The zero modes on the $S^2$ give rise to short multiplets of $\cN=(0,2)$ supersymmetry on the torus. There are three cases, depending on the shifted R-charge (\ref{shifted R def})~: 

\paragraph{Case $\rs=1$.} For $\rs=1$, there are no zero modes.

\paragraph{Case $\rs > 1$.} The fermionic zero modes for $\rs>1$ were given in (\ref{psi-exp}), with $j_0=\frac{\rs}{2}-1$ and $m=-j_0,\ldots,j_0$. They constitute a spin $j_0$ representation of $SO(3)$. The corresponding two-dimensional fermions $b_{j_0 m}$ are paired with the corresponding modes $f_{j_0 m}$ of $F$. The short multiplet $(b,f)$ realizes a  fermi multiplet \cite{Witten:1993yc} of the $\cN=(0,2)$ supersymmetry,
\bea\label{bf-zmode-var}
\delta b &= f~, &\qquad \t \delta b &= 0~,    \\
\delta f &=0~, &\qquad \t \delta f &=   2i D_{\bar{w}} b~.
\eea
These variations are simply a truncation of (\ref{S-var-csf}) for $\lambda=0$. The supersymmetric Lagrangian reads
\be  \label{bf-Lag}
\mathscr{L}_{(b,f)} = - \tilde{f}f + 2 i \tilde{b} D_{\bar{w}} b~.
\ee

\paragraph{Case $\rs < 1$.} For $\rs <1$, the fermionic zero modes were given in (\ref{psi-exp ii}), with $j_0=\frac{|\rs|}{2}$ and $m=-j_0,\ldots,j_0$. The corresponding two-dimensional fermions $c_{j_0 m}$ are paired with the $|\rs|+1$ bosonic zero modes $a_{j_0 m}$ of the scalar $\phi$.
The short multiplet $(a,c)$ is a chiral multiplet \cite{Witten:1993yc} of  $\cN=(0,2)$ supersymmetry,
\bea\label{ac-zmode-var}
\delta a &= c~, & \qquad \t \delta a &= 0~,  \\
\delta c &=0~, &\qquad \t \delta c &= 2iD_{\bar{w}} a~.  \\
\eea
Its supersymmetric Lagrangian is
\be \label{ac-Lag}
\mathscr{L}_{(a,c)} = 2D_{\bar{w}} \tilde{a} D_w  a + 2D_w \tilde{a} D_{\bar{w}} a + 2i \gamma(\tilde{a} D_{\bar{w}} a - a D_{\bar{w}} \tilde{a}) +2 i \tilde{c} D_{w} c + 2\gamma \tilde{c}c~.
\ee

\subsubsection{Twisted Boundary Conditions on the Torus}
We started with a theory on $T^2 \times S^2$ and reduced it to a theory on a torus with modular parameter $\tau$. At first sight, all dependence on the second complex structure parameter $\sigma= \tau\alpha -\beta$ has  disappeared from the effective theories \eqref{L abcf},\eqref{bf-Lag},\eqref{ac-Lag}. However, this dependence remains through twisted boundary conditions for the fields on the torus. Consider the expansion \eqref{expansion phi} of the scalar $\phi$. The monopole harmonics $Y_{\rs\, jm}$ satisfy  $Y_{\rs\, jm} \sim  e^{2\pi i (m + {\rs\over 2}) \alpha} Y_{\rs\, jm}$ for $x\sim x+ 2\pi$, and $Y_{\rs\, jm} \sim  e^{2\pi i (m + {\rs\over 2}) \beta} Y_{\rs\, jm}$ for $y\sim y+ 2\pi$. Since $\phi$ satisfies the boundary conditions \eqref{twisted bd ii}, each mode $a$ of angular momentum $j,m$ must satisfy
\bea\label{boundary condtions modes}
a(w+ 2\pi, \bw + 2\pi) &= e^{-2\pi i m \alpha} a(w,\bw)~, \cr
a(w+ 2\pi \tau, \bw + 2\pi \b\tau) &= e^{-2\pi i m \beta} a(w,\bw)~. 
\eea
The modes $b, c, f$ also satisfy these boundary conditions.~\footnote{Equivalently, we could choose periodic boundary conditions on $T^2$ and introduce a flat connection which couples to all the modes through their angular momentum $J_3=m$.}
The momentum operators along the one-cycles of the torus on modes $a,b,c,f$ of angular momentum $m$ are
\be
P_x = -i \d_x + m \alpha~, \qquad
P_y = -i \d_y+ m\beta~.
\ee
In the following, we will consider $y$ to be the ``time'' coordinate for canonical quantization while $x$ will be the space coordinate. Then  $P_x$ corresponds to the momentum denoted by $P$ in  (\ref{H short mult}), while $-i P_y$ corresponds to the Hamiltonian.

\subsection{Canonical Quantization and Supersymmetric Index}
Let us now compute the index (\ref{def S2T2 index}). To do so, we could further reduce the two-dimensional theories of the previous subsection on a circle, to obtain a quantum mechanics for states on $S^1\times S^2$. However, it will be more illuminating to directly quantize the theory on the torus.

The only states that contribute to the index sit in short multiplets (\ref{bf-zmode-var}) or (\ref{ac-zmode-var}). We will focus on those in the following. For $\rs=1$ there are no short multiplets, all states are paired by supersymmetry, and the index is simply
\be
\cI_{\rs=1}(q,x,t) = 1~.
\ee
Let us next consider the non-trivial cases $\rs \neq 1$. We choose the direction $y$ on the torus to be the time direction for canonical quantization. The boundary condition along the remaining spatial direction $x$ is given by the first line in \eqref{boundary condtions modes}.

\subsubsection{Case $\rs >1$~: The Fermi Multiplet}
The theory (\ref{bf-Lag}) is particularly simple. The auxiliary field  $f$ has no on-shell degrees of freedom and the equation of motion of $b$ is $D_\bw b=0$. It follows from \eqref{bf-zmode-var} that the supercharges annihilate all the states in this sector. Let us also note that the index for $\rs>1$ is trivially independent of the constant $\gamma$, since it does not appear in (\ref{bf-Lag}).

The fermionic wave functions are given by
\be
b = e^{i  \jq v_{\bar{w}} (\bar{w}-w)} e^{-i \alpha m w}\sum_{k=-\infty}^{\infty} \frac{e^{-ikw}}{\sqrt{2 \pi}} b_k~,  \qquad
\tilde{b} = e^{-i \jq v_{\bar{w}} (\bar{w}-w)} e^{i \alpha m w} \sum_{k=-\infty}^{\infty} \frac{e^{ikw}}{ \sqrt{2 \pi}} \tilde{b}_k~.
\ee
The canonical commutation relations imply 
\be
\{ b_k, \tilde{b}_l \} = \delta_{kl}~,
\ee
and the Hamiltonian is
\be \label{bf hamiltonian}
H_{(b,f)} = - i \sum_k \left(\tau k+ m \sigma + \jq \nu \right)\tilde{b}_k b_k~.
\ee
Recall that $\sigma= \tau \alpha - \beta$ and $\nu = \tau a_x - a_y$. We assume that the geometric moduli $\tau, \sigma$ and $\nu$ are given generic values, such that the Hamiltonian has no zero modes. The operators $b_k$ are either annihilation or creations operators, depending on $k$. Demanding that the real part of $H_{(b,f)}$ be positive, we find that $b_k$ is an annihilation operator for $k+ m\alpha + \jq a_x  > 0$, while it is a  creation operator for $k+ m\alpha + \jq a_x <0 $. Upon reordering $b_k$ and $\tilde b_k$ so that all modes have positive excitation energy, we obtain a Casimir energy~:
\bea\label{E0 ac mult}
E_0 &= i \sum_{k > m \alpha + q_f a_x} (k \tau - m \sigma - q_f \nu ) \cr
&= -i\tau \left(\frac{1}{12} + \frac{l_m(l_m+1)}{2} \right)+ i (m\sigma + q_f \nu) \left(\frac{1}{2} + l_m \right)~.
\eea
In the second line we used zeta function regularization,~\footnote{Here we used the Riemann zeta function after splitting the sum in \eqref{E0 ac mult}. Using the Hurwitz zeta function instead, we would find an additional term $-{i\over 2 \tau}(m\sigma + \jq \nu)^2$. We are assuming that this quadratic term is scheme dependent.} and $l_m$ is defined as the integer such that $l_m < m\alpha + q_f a_x < l_m+1$.

We now compute the index 
\be\label{trace bf sector}
\cI = \mathrm{Tr}\left((-1)^F e^{-2\pi H_{(b,f)}}\right)~.
\ee
Here the trace includes a sum over the $\rs-1$ copies of the $(b,f)$ multiplets (with $m= -{\rs\over 2}+1, \ldots, {\rs\over 2}-1$).
Let us define the fugacities
\begin{align}  \label{def fug}
q &= e^{2\pi i \tau}\, , & x &= e^{2\pi i\sigma}\, , & t &= e^{2\pi i \nu}~.
\end{align}
Using the explicit form of the Hamiltonian in \eqref{bf hamiltonian}, the single particle index is given by
\be\label{Isp}
\cI_{\mathrm{sp}} = -\sum_{m= -{\rs\over 2}+1}^{ {\rs\over 2}-1} \left( \sum_{k > m\alpha+\jq \alpha_x} q^k x^{-m} t^{-\jq} + \sum_{k > - m\alpha- \jq a_x} q^k x^{m} t^\jq \right)~,
\ee
The first sum over $k$ comes from the single particle states $b_k | 0\rangle$ with $k+ m \alpha + \jq a_x < 0$. They have quantum numbers $P=-k$, $J_3=-m$, $\JQ =-\jq$. It is convenient to change $k \rightarrow -k$ such that these states are weighted by $q^k x^{-m} t^{-q_f}$. Similarly, the second sum over $k$ in  \eqref{Isp} corresponds to states $\t b_k  | 0\rangle$ with $k+ m \alpha + \jq a_x>0$. 
The full index is obtained by  plethystic exponentiation \cite{Kinney:2005ej,Aharony:2003sx, Benvenuti:2006qr, Feng:2007ur}~:
\begin{align}\nn
 \prod_{m= -{\rs\over 2}+1}^{{\rs\over 2}-1} \left[ \frac{q^{1/12}}{\sqrt{x^m t^\jq}} \frac{q^{l_m(l_m+1)/2}}{(x^m t^\jq)^{l_m}} \prod_{k >  m \alpha+ \jq a_x} \left(1-  q^k x^{ -m } t^{-\jq} \right) \prod_{k >  - m \alpha- \jq a_x} \left(1-  q^k x^{ m } t^\jq \right) \right]~,
\end{align}
where we also included the contribution from the Casimir energy \eqref{E0 ac mult}.
Up to a constant phase which we discard, the term inside the square brackets can be written as
\begin{align}  \label{}
\frac{q^{1/12}}{\sqrt{x^m t^\jq}}\prod_{k \geq 0} \left(1-  q^{k+1} x^{ -m } t^{-q_f} \right)  \left(1-  q^k x^{ m } t^{q_f} \right) &= -i \frac{\theta_1(m\sigma + q_f \nu,\tau)}{\eta(\tau)}~.
\end{align}
Here we introduced the theta function \cite{Chandrasekharan}
\begin{align}  \label{def theta1}
\theta_1(\rho,\tau) &= 2 q^{\frac{1}{8}} \sin \pi\rho \prod_{k=1}^{\infty}(1-q^k)(1-q^ky)(1-q^k y^{-1})~,
\end{align}
with $y = e^{2\pi i \rho}$, and the eta function  $\eta(\tau)=q^{1/24} \prod_{k \geq 1}(1-q^k)$. The index for a four-dimensional $\cN=1$ chiral multiplet of shifted $R$-charge $\rs >1$ is then given by
\begin{align} \label{index rs g 1}
\cI_{\rs>1} &= \left( \frac{q^{1/12}}{\sqrt{ t^\jq}} \right)^{\rs -1 }\prod_{m= -{\rs\over 2}+1}^{{\rs\over 2}-1} \prod_{k \geq 0} \left(1-  q^{k+1} x^{ -m } t^{-q_f} \right)  \left(1-  q^k x^{ m } t^{q_f} \right) \nonumber \\
&= \prod_{m= -{\rs\over 2}+1}^{{\rs\over 2}-1} \frac{\theta_1(m\sigma + q_f \nu,\tau)}{i \, \eta(\tau)} \, . 
\end{align}
This is the result \eqref{result} advertised in the introduction. It also agrees with the result of \cite{Benini:2013nda,Benini:2013xpa}  for $\cN=(0,2)$ fermi multiplets.

\subsubsection{Case $\rs<1$~: The Chiral Multiplet}

Consider the $\cN=(0,2)$ chiral mutiplet with Lagrangian  \eqref{ac-Lag}. Unlike for the fermi multiplet, not every state in this sector contributes to the index. The equations of motion are
\be
D_{\bar{w}} (D_{w}  - i\gamma ) a = 0~,\qquad (D_{w} - i\gamma ) c = 0~,
\ee
and their most general solution takes the form
\be
a(w,\bar{w}) = e^{i \jq v_\bw \bw} a_{\mathrm{H}}(w) + e^{i(\jq v_w +\gamma) w}a_{\mathrm{AH}}(\bar{w})~,\qquad  c(w,\bw) = e^{i(\jq v_w+ \gamma) w}c_{\mathrm{AH}}(\bar{w})~.
\ee
It follows from \eqref{ac-zmode-var} that the anti-holomorphic modes $a_{\mathrm{AH}}, c_{\mathrm{AH}}$ are paired together and therefore do not contribute to the index. We thus  focus on the holomorphic modes of $a$. We have the mode expansion 
\bea
( D_{w} - i\gamma)a  &= i e^{i \jq v_{\bar{w}} (\bar{w}-w)} e^{-i \alpha m w} \sum_{k=-\infty}^{\infty} \frac{ e^{-ikw}}{\sqrt{2\pi}}a_k~,\cr
(D_{w} + i \gamma)\tilde{a} &= i e^{-i \jq v_{\bar{w}} (\bar{w}-w)}e^{i \alpha m w} \sum_{k=-\infty}^{\infty} \frac{e^{ikw}}{\sqrt{2\pi}} \tilde{a}_k~.
\eea
The canonical commutation relations are equivalent to 
\begin{align}  \label{}
[a_k,\tilde{a}_l] &= \frac{1}{2}(k+ m\alpha+ \jq a_x + \gamma)\delta_{k l}~.
\end{align}
Depending on the value of $k$, the operators $a_k$ are either creation or annihilation operators, like for the $(b,f)$ multiplet discussed above. For $k$ such that  $k + m\alpha+\jq a_x + \gamma > 0$,  $\tilde a_k$ is the  creation operators. When $k + m\alpha+\jq a_x + \gamma < 0$, we change notation $k \rightarrow -k$ and $a_{-k}$ is the creation operator. As before, we are assuming that the geometric parameters are generic.

The Hamiltonian acting on the ``holomorphic'' states (created by $\t a_k$ or $a_{k}$) reads
\be\label{holo H ac}
H_{a_{\mathrm H }} = -i \sum_k(k\tau + m \sigma + \jq \nu)\frac{2\tilde{a}_k a_k}{(k+ m\alpha + \jq a_x + \gamma)}~.
\ee
The complete Hamiltonian, including the contribution from the anti-holomorphic modes, is rather more complicated and will be omitted here.
The operators $a_k, \tilde a_k$ in \eqref{holo H ac} are not normal ordered. As in the $\rs >1$ case, this leads to a zero point energy which we compute by zeta function regularization.  

One can obtain the index for the four-dimensional chiral multiplet with $\rs <1$ by summing over the $|\rs|+1$ copies of the $(a,c)$ multiplet and using the same arguments as in the last subsection. One finds
\bea\label{result index rs less than 1}
\cI_{\rs<1} &= \left( \frac{\sqrt{t^\jq}}{q^{1/12}} \right)^{|\rs| + 1 }\prod_{m= -{|\rs|\over 2}}^{{|\rs|\over 2}} \prod_{k \geq 0} \frac{1}{\left(1-  q^{k+1} x^{ -m } t^{-q_f} \right)} \frac{1}{\left(1-  q^k x^{ m } t^{q_f} \right)} \cr
&= \prod_{m= -{|\rs|\over 2}}^{{|\rs|\over 2}} i \frac{ \eta(\tau)}{\theta_1(m\sigma + q_f \nu,\tau)}~. 
\eea
This gives \eqref{result ii}, and it agrees with the result of \cite{Benini:2013nda,Benini:2013xpa}  for $\cN=(0,2)$ chiral multiplets.
Note that the arbitrary parameter $\gamma$ does not contribute to the final answer.

\subsection{Modular Properties of $\cI^\Phi_{T^2\times S^2}$}\label{sec: mod prop}

Let us briefly comment on the modular properties of the $\cN=1$ chiral multiplet index $\cI^\Phi_{T^2\times S^2}$ obtained above. This partition function depends on three continuous parameters~: $\tau$, $\sigma$ and $\nu$. The first two parameters are the complex structures moduli of $T^2 \times S^2$ while the third is a holomorphic line bundle modulus for the $U(1)_f$ global symmetry. Those are the only continuous parameters on which the partition function is allowed to depend \cite{Closset:2013vra}. 

As explained in section 3, two different values of the parameters $\tau$ and $\sigma$ correspond to the same complex structure if they are related by the following transformations~:
\bea\label{Gcsi ii}
& S \; :  \; \; (\tau, \sigma) \mapsto \left(-{1\over \tau},\, {\sigma \over \tau}  \right)\, , \qquad\quad
& T \; :  \; \; (\tau, \sigma) \mapsto (\tau+1, \, \sigma)\, , \cr
& U \; :  \; \; (\tau, \sigma) \mapsto (\tau, \, \sigma +\tau )\, , \qquad\quad
& V \; : \; \; (\tau, \sigma) \mapsto (\tau, \,  \sigma + 1)\, ,
\eea
These operations generate a slight generalization of the modular group of the torus.
While  one may naively expect the partition function to be modular invariant, this is actually not the case.
Since these transformations correspond to large diffeomorphisms, the failure of modular invariance is interpreted as a gravitational anomaly. It would be interesting to understand this point better.

The modular group also acts on $\nu$ in the standard way, $S :  \nu \mapsto \nu/ \tau$ and $T :  \nu \mapsto \nu$.
Additionally, since the flat connections $a_x$ and $a_y$ are identified as $a_x\sim a_x + 1$, $a_y\sim a_y+1$ by large gauge transformations, we have the additional transformations $U' : \nu \mapsto \nu + \tau$ and $V' : \nu \mapsto \nu + 1$ which  are independent of $U$ and $V$. These large gauge transformations $U', V'$ are also anomalous.

Using the representation of the index in terms of the theta function \eqref{def theta1}, one can find its behaviour under the modular transformations described above. We have the following transformations~:
\begin{align}  \label{mod trans}
\frac{\theta_1(\rho, \tau+1)}{\eta(\tau+1)} &= e^{\pi i/6} \, \frac{\theta_1(\rho, \tau)}{\eta(\tau)}~, & 
\frac{\theta_1(\rho+\tau, \tau)}{\eta(\tau)} &= -e^{-i\pi \tau}e^{-2\pi i \rho} \, \frac{\theta_1(\rho, \tau)}{\eta(\tau)},  \nonumber \\
\frac{\theta_1(\rho+1, \tau)}{\eta(\tau)} &= -\frac{\theta_1(\rho, \tau)}{\eta(\tau)}~, & 
\frac{\theta_1\left(\frac{\rho}{\tau} , -\frac{1}{\tau} \right)}{\eta \left(-\frac{1}{\tau}\right)} &= -ie^{\frac{\pi i\rho^2}{\tau}} \, \frac{\theta_1(\rho, \tau)}{\eta(\tau)}~.
\end{align}
In this work, we will not adress the question of whether it is possible to restore some of the  modular properties by tuning some local terms. A better understanding of the scheme dependence of our answer would require a systematic understanding of the allowed supersymmetric counterterms, similar to \cite{Closset:2012vg,Closset:2012vp} in three dimensions. One can however see by inspection that it is not possible to achieve a fully modular invariant partition function.

\section{The Chiral Multiplet Partition Function on $\cM_4$}\label{sec: chiral mult Z}
In this section, we propose a simple method to compute the $\cN=1$ chiral multiplet partition function on any four-dimensional background with two supercharges. We evaluate the path integral explicitly using supersymmetric localization; see for instance \cite{Pestun:2007rz, Kapustin:2009kz, Jafferis:2010un, Hama:2010av, Hama:2011ea, Imamura:2011wg, Alday:2013lba} for similar arguments. 

After introducing some formalism in subsections \ref{subsec: frame}, \ref{subsec: chiral rev} and  \ref{subsec: inner product}, we will explain our path integral result in subsection \ref{subsec: eigenvalues}. In the  remaining  subsections we will apply our method to the case of $T^2\times S^2$ and $S^3\times S^1$.

\subsection{Building a Frame from the Killing Spinors}\label{subsec: frame}
Consider a background with two supercharges. From the Killing spinors $\zeta$ and $\t\zeta$, we can build the  vectors~\footnote{Note that $K, \b K$ and $Y, \b Y$ are not complex conjugates of each other. We hope that this choice of notation will not lead to any confusion.}
\be\label{defKXY}
K^\mu = \t\zeta \t\sigma^\mu \zeta~, \quad\quad  
\b K^\mu ={1\over 4}{ \t\zeta^\dagger \t\sigma^\mu \zeta^\dagger\over  |\zeta|^2  |\t \zeta|^2}~, \quad\quad  
Y^\mu =  {\t\zeta^\dagger \t \sigma^\mu \zeta \over 2 |\t \zeta|^2}~,
 \quad  \quad    
\b Y^\mu =- {\t\zeta\, \t \sigma^\mu \zeta^\dagger \over 2 |\zeta|^2  } ~.
\ee
Note that $Y$ and $\b Y$ have $R$-charge $\pm 2$, respectively. 
The vectors $K$ and $Y$ are anti-holomorphic with respect to the  complex structure ${J^\mu}_\nu$ defined in \eqref{csfromzeta}, while $\b K$ and $\b Y$ are holomorphic. Moreover, $Y$ and  $\b Y$ are valued in their respective $R$-symmetry line bundles $L^{\pm 2}$. The vectors \eqref{defKXY} are normalized such that
$K_\mu \b K^\mu = Y_\mu \b Y^\mu =\half$, and all other contractions of two of the vectors  \eqref{defKXY} vanish. These vectors provide a ($R$-charged) frame which will be very convenient below.

The vector $K$ is Killing, as already mentioned in section \ref{sec: susy with two supercharges}, while the three other vectors satisfy 
\be
\nabla_\mu \b K^\mu = 0~, \qquad D_\mu Y^\mu=0~, \qquad D_\mu \b Y^\mu=0~.
\ee
The covariant derivative $D_\mu$ was defined in \eqref{covder}. These relations follow from the Killing spinor equations \eqref{KSE}. 
It will be convenient to define the following operators acting on charged {\it scalars},
\be\label{def LKXY}
\hat\CL_K = K^\mu D_\mu~, \qquad \quad
\hat\CL_{\b K} = \b K^\mu D_\mu~, \qquad \quad
\hat\CL_Y = Y^\mu D_\mu~, \qquad \quad
\hat\CL_{\b Y} = {\b Y}^\mu D_\mu~.
\ee
 Note that  $\hat\CL_Y$ and $\hat\CL_{\b Y}$ shift the $R$-charge by $\pm 2$, respectively. The following lemma will be useful~:
\be\label{lemma CLK}
[\CLK, \CLKt]=0~, \qquad \quad  [\CLK ,\CLY]= 0~,\qquad\quad [\CLK, \CLYt]= 0~.
\ee
This lemma is most easily proven in the adapted coordinates $w,z$, in terms of which we have
\bea\label{XYZ with wz}
& K = \d_{\b w}~, &\qquad \quad  & Y ={ s\over c\, \Omega^2}\, ( \d_{\b z} - \b h \, \d_{\b w} )~, \cr
& \b K = {1\over \Omega^2} \d_w~,& \qquad \quad  & \b Y= {1\over c\, s}\, ( \d_{ z} -  h \, \d_{w} )~.
\eea
Here  the functions $\Omega(z,\b z)$, $h(z,\b z)$, $c(z,\b z)$ and $s(z,\b z)$ are the ones appearing in (\ref{2Q backgd i}),(\ref{2Q backgd ii}). 
Therefore, we see that \eqref{lemma CLK} is equivalent to
\be
 [D_\bw, D_\bz]= [D_w, D_\bw]=[D_z , D_\bw]=0~,
\ee
with the covariant derivative acting on scalar fields. For a field  of $R$-charge $r$ and $\JQ$ charge $\jq$, we have $[D_\mu, D_\nu]= -i r F_{\mu\nu}- i \jq f_{\mu\nu}$, where $F_{\mu\nu}$ is the field strength of the $R$-symmetry gauge field $A_\mu$ and $f_{\mu\nu}$ is defined in \eqref{def fmn for v}.  We already stated that $f_{\mu\nu}$ satisfies \eqref{background V}. We can similarly show that the Killing spinor equations imply
\be\label{constrain field strength FA}
 F_{\b w\b z}= F_{w\b w}=F_{z\b w} =0~,
\ee
in the presence of two Killing spinors $\zeta$, $\t\zeta$. (The first equation $F_{\bw\bz}=0$ follows from the existence of a single Killing spinor $\zeta$.) This proves \eqref{lemma CLK}. One can also compute
\be\label{YYt com}
[\CLY, \CLYt]= i (V^\mu -\kappa K^\mu) D_\mu +{r\over 8} (R-6 V^\mu V_\mu) - {\jq \over 2} D~.
\ee
Here $R$ is the Ricci scalar and $D$ the auxiliary background field  given by \eqref{background V}.

\subsection{The Chiral Multiplet Revisited}\label{subsec: chiral rev}
Consider a chiral multiplet coupled to an external gauge field, as discussed in section \ref{subsec: chiral mult}. 
For a chiral multiplet $\Phi=(\phi, \psi, F)$ of $R$-charge $r$ and $\JQ$-charge $\jq$, we can use the Killing spinor $\zeta$ to define
\be\label{defBC}
B= {1\over \sqrt2}{\zeta^\dagger \psi \over |\zeta|^2}~, \qquad C=\sqrt{2} \zeta\psi~, \qquad \Leftrightarrow	\qquad
\psi_\alpha = \sqrt2 \zeta_\alpha B-{1\over \sqrt2} {\zeta^\dagger_\alpha\over {|\zeta|^2} }C \, .
\ee
The fields $B$, $C$ are  anticommuting scalars of $R$-charge $r-2$, $r$, respectively. In terms of the variables $(\phi, B, C, F)$, the transformation rules (\ref{susychiralfdgauged}) read
\bea\label{susychiralBC}
\delta \phi &= C~, &\qquad \t \delta \phi &= 0~, \cr
\delta B &=  F~, &\qquad \t \delta B &=  -2 i \,\CLYt \phi~, \cr
\delta C &= 0~, &\qquad \t \delta C &= 2 i \, \CLK \phi~, \cr 
\delta F &= 0~, &\qquad \t \delta F &=2 i \left( \CLK B + \CLYt C \right)~.
\eea
For an anti-chiral multiplet $\t\Phi=(\t\phi,\t \psi,\t F)$ of charges $-r$ and $-\jq$, we use $\t\zeta$ to define
\be\label{defBCt}
 \t B= {1\over \sqrt2} {\t \zeta^\dagger \t\psi \over |\t \zeta|^2}~, \qquad   \t C= \sqrt2 \t \zeta\t \psi~, \qquad \Leftrightarrow	\qquad
\t \psi_\alphadot = \sqrt2  \t\zeta_\alphadot \t B - {1\over\sqrt2} {{\t\zeta^\dagger_\alphadot}\over {|\t\zeta|^2}} \t C \, .
\ee
The supersymmetry transformations (\ref{susyantichiralfdgauged}) become
\bea\label{susyachiralBC}
\delta \t \phi &= 0~, &\qquad \t \delta \t \phi &=  \t C~, \cr
\delta \t B &= 2 i \,\CLY \t \phi~, &\qquad \t \delta \t B &=  \t F~, \cr
\delta \t C &= 2 i \, \CLK \t \phi~, &\qquad \t \delta \t C &= 0~, \cr 
\delta \t F &= 2 i \left( \CLK \t B - \CLY \t C \right)~, &\qquad \t \delta \t F &=0~.
\eea
Using (\ref{lemma CLK}), one can see that the transformations  (\ref{susychiralBC}), (\ref{susyachiralBC}) realize the supersymmetry algebra $\{ \delta, \t \delta\} = 2i \CLK$. It should be emphasized that in the new variables the supercharges are scalars and $R$ neutral. 

It is illuminating to write the chiral multiplet Lagrangian \eqref{Lag Phi} in terms of the new variables. With the help of \eqref{YYt com}, one finds
\begin{align}  \label{twLag}
\mathscr{L}_{\Phi \t \Phi} &= 4\CLKt \t \phi \CLK \phi + 4 \CLY \t \phi \CLYt \phi + i \kappa (\CLK \t \phi \phi - \t \phi \CLK \phi) - \t F F \cr
& \quad  +2 i \t B \CLK B + 2 i \t C \CLKt C +2i \t B \CLYt C  - 2i \t C \CLY B  - \kappa \t C C. 
\end{align}
The results \eqref{susychiralBC}, \eqref{susyachiralBC}, \eqref{twLag} should be compared to \eqref{S-var-csf}, \eqref{L abcf}. We will see in the following that we can take the analogy to the $T^2 \times S^2$ computation further. For any background $\cM_4$ with two supercharges, only modes in shortened multiplets will contribute to the $\cN=1$ chiral multiplet partition function $Z^\Phi_{\cM_4}$ ---those short multiplets are of the form $(B, F)$ or $(\phi, C)$, similarly to the fermi and chiral $\cN=(0,2)$ multiplets of section \ref{sec: compute index}.

\subsection{Reality Condition and Inner Product}\label{subsec: inner product}
We would like to evaluate the path integral of a chiral multiplet with Lagrangian \eqref{twLag} explicitly.
In order to define the Euclidian path integral, we need to choose  a contour of integration. We will assign the following reality conditions to  the fields $\phi, B, C, F$ in a chiral multiplet of $R$-charge $r$~:
\be\label{real cond}
\t\phi = {\Omega^r \over |s|^r} \phi^\dagger~,\qquad\quad
\t C = {\Omega^r \over |s|^r} C^\dagger~,\qquad\quad
\t B = {\Omega^{r-2} \over |s|^{r-2}} B^\dagger~,\qquad\quad
\t F= -{\Omega^{r-2} \over |s|^{r-2}} F^\dagger~,
\ee
with $\Omega$ and $s$ the quantities defined in section \ref{sec: susy with two supercharges}. This prescription  is consistent with the complexified $R$-symmetry transformations and dimensional analysis (recall that $s$ has $R$-charge $2$).
It will also be useful to consider appropriate inner products on field space.
For the fields $\phi$ and $C$ of $R$-charge $r$, let us define the  inner product
\be\label{in prod 1}
\langle \phi_1, \phi_2 \rangle_r = \int \, d^4x \sqrt{g} \,{\Omega^{r}\over |s|^r} {\phi^\dagger_1\, \phi_2 \over \Omega^2}~. 
\ee
Similarly, for the fields $B$ and $F$ of $R$-charge $r-2$, we define
\be\label{in prod 2}
 \langle B_1, B_2 \rangle_{r-2} = \int \, d^4x \sqrt{g} \,{\Omega^{r-2} \over |s|^{r-2}}B^\dagger_1\, B_2~.  
\ee
We will denote by $\cH_r$ and $\cH_{r-2}$ the corresponding Hilbert spaces. Note that the operator $i \CLYt$ acting on $\phi$ or $C$ and the operator $i \Omega^2 \CLY$ acting on $B$ or  $F$ are maps between different Hilbert spaces,
\bea\label{adj op LY LYt}
i \CLYt~: \cH_r \rightarrow \cH_{r-2}~, \qquad\qquad i\Omega^2 \CLY~: \cH_{r-2} \rightarrow \cH_{r}~,
\eea
while $i\CL_K$ and $i \Omega^2 \CLKt$ act inside $\cH_r$ or $\cH_{r-2}$ (depending on the $R$-charge).
Moreover, the two operators \eqref{adj op LY LYt} are mutually adjoint (see Appendix \ref{App: Pairing})~:
\bea\label{adjoint property}
 &\langle \phi, i \Omega^2 \CLY B \rangle_{r}=  \langle i \CLYt \phi,  B \rangle_{r-2}~, \cr
& \langle B, i \CLYt \phi \rangle_{r-2}=  \langle i\Omega^2 \CLY B, \phi \rangle_{r}~.
\eea

\subsection{Localization and Unpaired Eigenvalues}\label{subsec: eigenvalues}
Let us consider the following $\delta$-exact  deformation of the $\cN=1$ chiral multiplet theory \eqref{twLag}~:
\be\label{Sloctop}
\mathscr{L}_{\rm loc}=  \delta\left( -2 i \t B \CLYt \phi - 2 i \t C \CLKt \phi - \t F B \right)~.
\ee
This term is equal to \eqref{twLag} itself with $\kappa =0$.  (Note that the full \eqref{twLag} is also $\delta$-exact.) It can be written
\be\label{Sloctopiii}
\mathscr{L}_{\rm loc}= 4\, \t\phi \big( - \CLKt \CLK - \CLY \CLYt\big) \phi +2 i\, \t\Psi \mat{ \CLK & \CLYt  \cr -\CLY  & \CLKt } \Psi \ - \t F F~,
\ee
where we introduced $\Psi = (B,C)^{T}$ and  $\t\Psi = (\t B, \t C)$. Let us define the kinetic operators
\be\label{kinops new}
\Delta_{\rm bos} = \Omega^2\big(- \CLKt \CLK - \CLY \CLYt\big)~, \qquad \quad \Delta_{\rm fer} =   i \,   \mat{ \CLK & \CLYt  \cr -\Omega^2\CLY  & \Omega^2 \CLKt }~.
\ee
Note the appearance of factors of $\Omega^2$ in this definition, necessary to make these operators dimensionless. Using \eqref{real cond} and \eqref{kinops new}, we can write the localizing action  in term of the inner products \eqref{in prod 1} and \eqref{in prod 2} as
\be\label{S loc inner prod}
S_{\rm loc} = \int d^4 x \sqrt{g}\, \mathscr{L}_{\rm loc} 
= 4\langle \phi, \Delta_{\rm bos} \phi \rangle_r + 2 \langle \Psi, \Delta_{\rm fer} \Psi \rangle_{r, r-2} + \langle F, F \rangle_{r-2}~,
\ee
where we defined
\be
 \langle \Psi_1 ,  \Psi_2 \rangle_{r, r-2}=  \langle B_1, B_2 \rangle_{r-2}+ \langle C_1, C_2 \rangle_{r}~.
\ee
By a standard supersymmetric localization argument, we can deform the original theory by \eqref{S loc inner prod} with an arbitrarily large coefficient without affecting the path integral, which therefore reduces to a ratio of functional determinants,
\be\label{def Zloc}
Z_{\cM_4}^\Phi = {\det{\Delta_{\rm fer}} \over\det{\Delta_{\rm bos}}}~.
\ee
In deriving \eqref{def Zloc}, we are assuming that $\phi=0$ is the only relevant saddle point. (This is true for generic values of the complex structure and line bundle moduli.)
This result can be simplified further. Let us first note that
\be\label{rel dets}
 i \,   \mat{ \CLK & \CLYt  \cr -\Omega^2\CLY  & \Omega^2 \CLKt }\mat{1 & - i  \CLYt \cr 0 & i \CLK } = \mat{i\CLK & 0 \cr -i  \Omega^2\CLY & \Delta_{\rm bos}}~,
\ee
where we used  $[\CLK,\CLYt]=0$. Taking the determinant on both sides of \eqref{rel dets}, we find 
\begin{align}  \label{SUSYdet}
\frac{\det \Delta_{\rm fer} } { \det \Delta_{\rm bos} } &= \frac{\det (i\CLK^{(r-2)})}{\det (i\CLK^{(r)})}~.
\end{align}
Here we denoted by $i\CLK^{(r-2)}$ and $i\CLK^{(r)}$  the operator $i\CLK$ acting on $\cH_{r-2}$ and $\cH_{r}$, respectively.
Consider next the adjoint operators \eqref{adj op LY LYt}. A standard argument shows that 
\begin{align}  \label{}
{\rm Ker} (i\Omega^2 \CLY) &= {\rm Ker }(\CLYt \Omega^2\CLY)~, & {\rm Ker} (i\CLYt) &={ \rm Ker} (\Omega^2 \CLY \CLYt)~, 
\end{align}
and that the non-zero eigenvalues of $ \CLYt \Omega^2\CLY$ and $\Omega^2 \CLY \CLYt$ are in one-to-one correspondence. 
In other words, $i \Omega^2\CLY$ provides an isomorphism between the space of eigenfunctions of $\CLYt \Omega^2 \CLY$ in $\cH_{r-2}$ with non-vanishing eigenvalues and the space of eigenfunctions of $\Omega^2 \CLY \CLYt$ in $\cH_{r}$ with non-vanishing eigenvalues.
Additionally, by \eqref{lemma CLK}  this isomorphism commutes with $i\CLK$.
It follows that all the eigenvalues of $i\CLK$ which lie outside of ${\rm Ker} (i\Omega^2 \CLY)$ or ${\rm Ker} (i\CLYt)$ cancel from \eqref{SUSYdet}. Therefore, we find that
\be\label{Z loc result}
Z_{\cM_4}^\Phi=\frac{{\det}_{\rm Ker \CLY} (i\CLK^{(r-2)})}{ {\det}_{\rm Ker \CLYt} (i\CLK^{(r)})}~.
\ee
Similar arguments appeared, for instance, in \cite{Pestun:2007rz,  Hama:2011ea, Alday:2013lba}.
Practically speaking, the partition function reduces to
\be\label{Zloc eigen}
Z_{\cM_4}^\Phi =  {\prod \lambda_B \over \prod \lambda_\phi}~,
\ee
with the eigenvalues $\lambda_B$, $\lambda_\phi$ determined by the following BPS-like linear equations~:
\bea\label{eigenvaluesBPhi}
&i \CLK B = \lambda_B\,  B\, , \qquad \quad & \CLY B =0~, \cr
&i \CLK \phi = \lambda_\phi\, \phi\, , \qquad \quad &\CLYt \phi =0~.
\eea
It is also useful to rewrite these equations in terms of the complex coordinates $w, z$ using \eqref{XYZ with wz}~:
\bea\label{eigenvalueszw}
& i D_{\b w} B = \lambda_B\, B\, , \qquad \quad & (D_{\b z} - \b h \, D_{\b w})B=0 \, ,\cr
& i D_{\b w} \phi = \lambda_\phi\, \phi\, , \qquad \quad\quad & (D_{z} -h D_{ w})\phi =0 \, .
\eea

Let us note that the equation  $\CLY B =0$ in \eqref{eigenvaluesBPhi} is a condition for the shortening of the $\cN=1$ chiral multiplet. For a solution of this equation, it is consistent to set $C=0$ (and also $\phi=0$) because $B$ disappears from its equation of motion. We are then left with a short multiplet $(B,F)$ akin to the $(b,f)$ multiplet we found for $T^2 \times S^2$ in section \ref{sec: compute index}.
 Similar remarks apply to $(\phi,C)$ and the $(a,c)$ multiplet of our $T^2 \times S^2$ index computation. 

The equations \eqref{eigenvalueszw} are also consistent with known constraints on the parameter dependence of  $\cN=1$ supersymmetric partition functions \cite{Closset:2013vra}. We will see in the examples that the eigenvalues $\lambda_B, \lambda_\phi$ depend holomorphically on complex structure moduli and line bundle moduli  (up to an arbitrary overall rescaling). Moreover, for any background with two supercharges the equations \eqref{eigenvalueszw} are independent of the ambiguity $\kappa$ in \eqref{2Q backgd ii}.
It is indeed expected that the partition function of a chiral multiplet be independent of $\kappa$, because this free theory possesses an FZ multiplet \cite{Closset:2013vra}.

In the rest of this section, we will compute \eqref{Zloc eigen} in two interesting examples. For $\cM_4= T^2 \times S^2$, we will reproduce the canonical quantization result of section \ref{sec: compute index}. For $\cM_4=S^3\times S^1$, we will reproduce the previously known result \cite{Romelsberger:2007ec, Dolan:2008qi} in a new and elegant manner.

\subsection{The Case of $T^2\times S^2$} 
Let us compute the one-loop determinant \eqref{Zloc eigen} for the $T^2 \times S^2$ background of section \ref{complex_st}. Consider the $B$ modes first. They are solutions of the first line in \eqref{eigenvalueszw}, which becomes
\be\label{eigen eq num T2S2}
i \left(\d_\bw -i \jq v_\bw\right) B= \lambda_B \, B~, \quad\qquad \left(\d_\bz +{\rs -2\over 2}{z\over 1+|z|^2} \right) B=0~.
\ee
Here $\rs$ is the shifted $R$-charge and $v_\bw$ the $U(1)_f$ flat connection, as  defined in section \ref{subsec: backgd V}.
The second equation in \eqref{eigen eq num T2S2} implies $B = f_1(z) (1+|z|^2)^{-{\rs-2\over 2}} f_2(w,\bw)$, with $f_1(z)$ holomorphic. It is convenient to consider eigenmodes of the angular momentum operator $J_3 = z\partial_z - \bar z \partial_{\bar z} - \frac{\rs-2}{2}$ on the sphere (see Appendix \ref{App: monopole harmonics}). $J_3 B = m B$ implies $f_1\propto z^{m+ {\rs-2\over 2}}$. 
 We also consider definite momentum $n_x, n_y\in \Z$ on the torus. Taking into account the boundary conditions \eqref{twisted bd ii}, we find 
\be\label{sol B T2S2}
B= e^{\frac{1}{2\tau_2}(n_x \tau -n_y - m\sigma) \bw - \frac{1}{2\tau_2}(n_x \bar \tau -n_y - m\bar \sigma) w}\frac{z^{m + \frac{\rs-2}{2}}}{(1+z\bar z)^{\frac{\rs-2}{2}}}~.
\ee
Moreover, $B$ is normalizable if and only if $\frac{2-\rs}{2} \leq m \leq \frac{\rs-2}{2}$. In particular, solutions exist only if $\rs \geq 2$.
We easily see that \eqref{sol B T2S2} solves \eqref{eigen eq num T2S2} with eigenvalue
\be\label{eigenvalues B T2S2}
\lambda_B = \frac{i}{2 \tau_2}(n_x \tau -n_y - m\sigma - \jq \nu)~.
\ee
Note that the $B$ modes correspond to the first kind of spinor zero modes in  \eqref{fer-0-modes}.
A similar analysis for the modes $\phi$ solving \eqref{eigenvalueszw} leads to 
\be
\phi = e^{\frac{1}{2\tau_2}(n_x \tau -n_y - m\sigma) \bw - \frac{1}{2\tau_2}(n_x \bar \tau -n_y - m\bar \sigma) w} {(1+z\bar z)^{{\rs\over 2}}\over \bz^{m+{\rs \over 2}}}~,
\ee
with $\frac{\rs}{2} \leq m \leq -\frac{\rs}{2}$. In particular such solutions exist only for $\rs \leq 0$. The eigenvalues $\lambda_\phi$ are given by
\be\label{eigenvalues phi T2S2}
\lambda_\phi = \frac{i}{2 \tau_2}(n_x \tau -n_y - m\sigma - \jq \nu)~.
\ee
These modes correspond to the second set of spinor zero modes in \eqref{fer-0-modes}.

Therefore, we have three distinct cases, depending on the shifted $R$-charge $\rs \in \Z$. If $\rs=1$, all the bosonic and fermionic modes are paired together by supersymmetry and therefore the partition function is 
\be\label{Z one}
Z^\Phi_{T^2 \times S^2; \, \rs=1} = 1\, .
\ee
If  $\rs>1$, only the $B$ modes contribute to the partition function, with eigenvalues \eqref{eigenvalues B T2S2}~:
\be\label{Z r g one}
Z^\Phi_{T^2 \times S^2; \, \rs > 1} = \prod_{m = -{\rs\over 2}+1 }^{{\rs\over 2}-1}   \prod_{n_x, n_y = -\infty}^\infty  \big(  \tau n_x - n_y - m\sigma  - \jq \nu \big)~.
\ee
Here we rescaled away the overall factor of ${i\over 2 \tau_2}$  in \eqref{eigenvalues B T2S2}.~\footnote{Such an overall rescaling is arbitrary and does not affect the final answer in terms of $\zeta$-function regularized products.}
If $\rs<1$, the $\phi$ mode eigenvalues  \eqref{eigenvalues phi T2S2} are the ones which contribute~:
\be\label{Z r l one}
Z^\Phi_{T^2 \times S^2; \, \rs < 1} = \prod_{m = -{|\rs|\over 2} }^{|\rs|\over 2}   \prod_{n_x, n_y = -\infty}^\infty  {1 \over   \tau n_x - n_y - m\sigma  - \jq \nu  }~.
\ee
The infinite products (\ref{Z r g one}), (\ref{Z r l one}) need to be properly regularized, which can be done using zeta function regularizaton ---see in particular Example 13 of \cite{QuineZeta}. This leads to  \eqref{result}, \eqref{result ii}.

\subsection{The Case of $S^3\times S^1$~: The Elliptic Gamma Function}\label{subsec: S3 S1 Z}
The supersymmetric partition function of $\cN=1$ theories on $S^3\times S^1$, $Z_{S^3\times S^1}$, has been well-studied in the literature. It computes a supersymmetric index for the theory quantized on $S^3$ \cite{Kinney:2005ej,Romelsberger:2005eg}, which has been computed exactly for rather general $\cN=1$ gauge theories \cite{Romelsberger:2007ec, Dolan:2008qi}. In particular, the chiral multiplet partition function is given by a certain elliptic gamma function \cite{Dolan:2008qi}. In the following, we will rederive that last result using the method of section \ref{subsec: eigenvalues}.

Let us consider a complex manifold $\cM_4^{p,q}$ diffeomorphic to $S^3 \times S^1$ called a primary Hopf surface of the first type. It is obtained as a quotient of $\C^2-(0,0)$~:
\be\label{Hopf surface}
(z_1, z_2)\sim (p\, z_1, q\, z_2)~, \qquad 0< |p|\leq |q|<1~.
\ee
Here $z_1, z_2$ are the coordinates on $\C^2-(0,0)$, and $p, q$ are complex structure parameters. It was realized recently that the partition function $Z_{S^3\times S^1}$ is a locally holomorphic function on the complex structure moduli space of $\cM_4^{p,q}$  \cite{Closset:2013vra}. We refer to  \cite{Closset:2013vra} for more details and references.~\footnote{Note that our coordinates $z_1, z_2$ are denoted by $w,z$ in \cite{Closset:2013vra}. In the present paper we reserve the notation $w,z$ for the special coordinates adapted to two supercharges, such that $K= \d_\bw$ is the Killing vector built from the two Killing spinors.}

It will be convenient to introduce two complex parameters $\sigma=\sigma_1+ i\sigma_2$ and  $\tau=\tau_1+ i \tau_2$ (with $\sigma_{1,2}, \tau_{1,2}$ real) such that
\be\label{p q to sigma tau}
p=e^{2\pi i \sigma}~, \qquad q= e^{2\pi i \tau}~, \qquad 0<\tau_2\leq \sigma_2~, \tau_1 \sim \tau_1+1~, \sigma_1 \sim \sigma_1 +1~.
\ee
For generic values of $p,q$, $\cM_4^{p,q}$ admits  two supercharges \cite{Closset:2013vra}. In the rest of this subsection, we will consider the subcase $\sigma_2=\tau_2$ for simplicity. We consider the following Hermitian metric on $\cM_4^{p,q}$ (with $|p|=|q|$)~:
\be\label{metric z1z2}
ds^2 = {1\over |z_1|^{2} + |z_2|^{2}}\left( dz_1 d\bz_1 +  dz_2 d\bz_2 \right)~.
\ee
To take advantage of the general discussion of section \ref{sec: susy with two supercharges}, we introduce new coordinates
\be\label{coord wz i}
w= - i  \log z_1~, \qquad z= {z_2\over z_1}~.
\ee
These coordinates cover $S^3 \times S^1$ except for the locus $z_1=0$, which can be covered with coordinates $w'= w- i \log z$, $z'= {1\over z}$. The $w, z$ coordinates \eqref{coord wz i} are subject to the following identifications~:
\be\label{periodicities S3S1 wz}
\big(w,z\big)\sim  \big(w+ 2\pi {\sigma} ,\, e^{2\pi i (\tau_1 -\sigma_1)} z\big)~,\qquad
\big(w,z\big)\sim  \big(w+ 2\pi  ,\,  z\big)~,
\ee
It is convenient to consider real coordinates $x, \theta, \varphi, \chi$ on $S^3\times S^1$, with $x\in [0, 2\pi )$ the coordinate on the $S^1$ and $\theta\in [0,\pi]$, $\varphi, \chi \in [0, 2\pi)$ the coordinates on the $S^3$. In terms of these angular coordinates,
\be
w= \sigma x +\varphi - i\log{\left(\cos{\theta\over 2}\right)}~, \qquad z= e^{i(\tau_1 -\sigma_1) x}\tan{\theta\over 2} e^{i(\chi-\varphi)}~.
\ee
The periodicities \eqref{periodicities S3S1 wz} correspond to $x\sim x+ 2\pi$ and $\varphi \sim \varphi + 2\pi$, respectively.
In terms of the $w, z$ coordinates, the metric  \eqref{metric z1z2} takes the  form  
\be\label{metric S3S1 wz}
ds^2=\left(dw - {i\bz\over  1+|z|^2}  d z\right)\left(d\b w + {iz\over  1+|z|^2}  d \b z\right) + {1 \over  (1+|z|^2)^2} dz d\b z~.
\ee
The resulting supersymmetric background is studied in more detail in Appendix  \ref{App: S3S1}. In addition to the supergravity background fields, we also consider a background gauge field for a $U(1)_f$ internal symmetry,
\be\label{bckgd gauge field S3S1}
a_\mu dx^\mu = {1\over 2 i \sigma_2}(a_r + i a_i)\left(dw - {i\bz\over  1+|z|^2}  d z\right)  - {1\over 2 i \sigma_2}(a_r - i a_i)\left(d\b w + {iz\over  1+|z|^2}  d \b z\right)~,
\ee
which preserves the same supercharges as \eqref{metric S3S1 wz}. Let us also define the fugacity
\be\label{def u S3S1}
u= e^{-2 \pi i(a_r-i a_i)}~.
\ee
The complex parameter $a_r- i a_i$ is a  holomorphic line bundle modulus \cite{Closset:2013vra}.

For generic values of $\tau_1, \sigma_1$ and $\sigma_2= \tau_2$, the background \eqref{metric S3S1 wz} has a $U(1)^3$ isometry, corresponding to rotations along the real angles $x, \varphi, \chi$ mentioned above.  Supersymmetry dictates twisted periodicities for $R$-charged fields going around these angles.
 The corresponding momentum operators acting on fields of $R$-charge $r$ are given by
\bea\label{P ops on S3S1}
P_x &= (\tau_1-\sigma_1) (z\d_z -\bz \d_\bz )  -i(\sigma \d_w +\b\sigma \d_\bw) -{r\over 2}(\tau_1-\sigma_1) ~,\cr
P_\varphi &=-  (z\d_z -\bz \d_\bz ) - i ( \d_w + \d_\bw) +r ~,\cr
P_\chi &=  (z\d_z -\bz \d_\bz )~.
\eea
These operators have integer eigenvalues.  Note that 
\be\label{K from Ps}
K =\d_\bw= {1\over 2 \sigma_2} \left(-P_x +\sigma P_\varphi  + \tau P_\chi  -{r\over 2} (\sigma+ \tau) \right)~.
\ee

Let us compute the partition function for a chiral multiplet of $R$-charge $r$ and $\JQ$-charge $\jq$, according to \eqref{Zloc eigen}, \eqref{eigenvalueszw}.
Using the results of Appendix  \ref{App: S3S1}, the eigenvalue equations for the unpaired modes of type $B$ read
\be\label{eq for B S3S1}
 i (\d_\bw - i \jq v_\bw ) B =\lambda_B\, B~, \qquad \left( \d_\bz + {r-2\over 2} {z\over 1+|z|^2} - {iz\over 1+|z|^2}\d_\bw \right) B=0~,
\ee
where $v_\bw =  - {1\over 2 i \sigma_2}(a_r - i a_i)$. We also require $B$ to be an eigenmode of the operators \eqref{P ops on S3S1}, with 
\be
P_x B = n_0 B~, \qquad P_\varphi B= n_1 B~, \qquad P_\chi B = n_2 B~, \qquad n_0, n_1, n_2 \in \Z~.
\ee
Therefore, the first equation in \eqref{eq for B S3S1} together with \eqref{K from Ps} imply
\be\label{lamB S3S1}
\lambda_B = {i\over 2  \sigma_2} \big(-n_0 + n_1\sigma + n_2 \tau -{r-2\over 2}(\sigma+ \tau) + \jq (a_r - ia_i)  \big)~.
\ee
The $B$ modes  are given explicitly by
\be
B = z^{n_2 } (1+|z|^2)^{\lambda_B - \jq v_\bw - {r-2\over 2}}\, e^{-i (\lambda_B - \jq v_\bw) \bw} e^{-i (\b\lambda_B - \jq v_w) w}~.
\ee
A similar expression holds in the $w',z'$ patch, with  $|B| \sim |z'|^{n_1}$ near $z' =0$.
Normalizabiltiy of the modes restricts the allowed values of $n_1, n_2$ to  $n_1, n_2 \geq 0$.
Similarly, the  $\phi$ modes solving  \eqref{eigenvalueszw} are given by
\be
\phi= {1\over \bz^{n_2}} (1+ |z|^2)^{-(\b\lambda_\phi - \jq v_w -{r\over 2})}\, 
 e^{-i (\lambda_\phi - \jq v_\bw) \bw} e^{-i (\b\lambda_\phi - \jq v_w) w}~.
\ee
with eigenvalues
\be\label{lamphi S3S1}
\lambda_\phi = {i\over 2 \sigma_2} \big(-n_0 + n_1\sigma + n_2 \tau -{r\over 2}(\sigma+ \tau) + \jq (a_r - ia_i)  \big)~.
\ee
The integers $n_0, n_1, n_2$ are again the eigenvalues of \eqref{P ops on S3S1}.
Nomalizability imposes $n_1, n_2 \leq 0$.
Plugging the eigenvalues \eqref{lamB S3S1},\eqref{lamphi S3S1} into \eqref{Zloc eigen} and renaming some of the integers, we find the partition function
\be\label{Z S3S1 inf prod}
Z^\Phi_{S^3\times S^1} = 
 \prod_{n_0 = -\infty}^\infty \prod_{n_1, n_2 = 0}^\infty   { n_0+ \sigma n_1 +\tau n_2 - {r-2\over 2} (\sigma+\tau) +  \jq (a_r- i a_i) \over  n_0+ \sigma n_1 +\tau n_2 + {r\over 2} (\sigma+\tau) - \jq (a_r-i a_i)}~.
\ee
Note that this formula is a natural generalization of the three-dimensional localization result for the squashed $S^3$ \cite{Hama:2011ea,Imamura:2011wg}.

The result \eqref{Z S3S1 inf prod} can be regularized using  Barnes' multiple zeta function \cite{Ruijs,FriedmanRuijSB}.~\footnote{One can rewrite \eqref{Z S3S1 inf prod} in terms of triple gamma functions and use Corollary 6.2 of \cite{FriedmanRuijSB}.}  In term of the parameters $p,q, u$ defined in \eqref{p q to sigma tau} and \eqref{def u S3S1}, we obtain
\bea\label{result S3S1}
Z^\Phi_{S^3\times S^1}(p, q, u)&=e^{i \pi  \cA}  \prod_{j,k =0}^\infty {1 - u^{-\jq} p^{j+1-{r\over 2}} q^{k+1-{r\over 2}}\over  1 - u^\jq p^{j+{r\over 2}} q^{k+{r\over 2}}  }\cr
&=  e^{i  \pi   \cA} \, \Gamma_e (u^\jq (pq)^{r\over 2}; p, q)~.
\eea
The function $\Gamma_e(t; p, q)$ is the elliptic gamma function.~\footnote{
It can be defined by
\be\nonumber
\Gamma_e(t; p, q) = \prod_{j,k =0}^\infty {1 - t^{-1} p^{j+1} q^{k+1}\over  1 - t p^j q^k  }~. 
\ee}
We thus reproduced the known result for the $\cN=1$ chiral multiplet $S^3\times S^1$ partition function, without relying on the supersymmetric index point of view of \cite{Romelsberger:2007ec, Dolan:2008qi, Gerchkovitz:2013zra}.~\footnote{See  \cite{Nawata:2011un} for a related computation of the $\cN=4$ index using localization.} The answer is locally holomorphic in the geometric moduli $\tau, \sigma$ and $a_r - i a_i$, as expected \cite{Closset:2013vra}. In addition to the expected gamma function, there is an interesting prefactor $e^{i \pi  \cA}$ in \eqref{result S3S1}, with $\cA$ given by
\be
\cA({\bf w}) = {1\over \sigma \tau} \left({\bf w}^3 - {\tau^2 + \sigma^2 -2\over 12}{\bf w} \right)~, 
\qquad \quad {\bf w} = (r-1){\tau +\sigma\over 2} - \jq (a_r - i a_i)~.
\ee
Note that $\cA$ is a cubic polynomial in $r-1$ and $\jq$, which are the $R$- and $\JQ$-charges of the fermion in the $\cN=1$ chiral multiplet.  It would be interesting to understand whether this $\cA$ is physical or whether it can be removed by a local counterterm.

\subsection{Computation with Arbitrary Hermitian Metrics}
In the  two examples above, we chose rather simple Hermitian metrics for ease of presentation. However, it is easy to generalize the computation of section \ref{subsec: eigenvalues} to arbitrary backgrounds with two supercharges. Consider for instance the case of $T^2 \times S^2$ with an arbitrary Hermitian metric compatible with a given complex structure (with parameters $\tau,\sigma$). Using the $w, z$ coordinates, we can solve the eigenvalue equations \eqref{eigenvaluesBPhi} with the ansatz
\bea
B &= e^{i(n_x +{\rs-2\over 2}\alpha  ) x}  e^{i (n_y +{\rs-2\over 2} \beta)y }e^{i (m+{\rs-2\over 2}  ) \varphi} B_0(\theta)~,\cr
\phi &=e^{i(n_x +{\rs\over 2}\alpha  ) x}  e^{i (n_y +{\rs\over 2} \beta)y }e^{i (m+{\rs\over 2}  ) \varphi} \phi_0(\theta)~,
\eea
in the northern patch. Here $x, y, \varphi, \theta$ are the real coordinates introduced in \eqref{cc on S2T2}. This ansatz is dictated by the periodicities \eqref{twisted bd ii} (and by the general properties of the spin operator $J_3$ on $S^2$ with magnetic flux), which are independent of the details of the metric (although they do depend on the choice of complex coordinates).  Using the expression \eqref{K real coord} for $\d_\bw$, we directly find the eigenvalues $\lambda_B$ \eqref{eigenvalues B T2S2} and  $\lambda_\phi$ \eqref{eigenvalues phi T2S2}. Possible restrictions on the values of the integers in $\lambda_{B,\phi}$ can be obtained from a careful analysis of the profiles $B_0(\theta), \phi_0(\theta)$ near $\theta=0, \pi$, asking that the modes be normalizable. (Understanding the limit $\theta\sim 0, \pi$ is much simpler than solving for the full profiles, which cannot be done in general.)

A similar analysis can be done for $S^3 \times S^1$. One thus confirms explicitly, in our simple examples,  that the partition function on $\cM_4$ is independent of the Hermitian metric \cite{Closset:2013vra}.

\section*{Acknowledgments}
\label{s:acks}

It is a pleasure to thank Ofer Aharony, Francesco Benini, Lorenzo Di Pietro, Thomas Dumitrescu, Guido Festuccia, Itamar Yaakov and especially Zohar Komargodski  for stimulating discussions and  for comments on the manuscript.  We also thank Joel Fine for pointing out reference \cite{Atiyah:1955}. 
I.S. would like to extend his gratitude to Efrat, Guy, Nizan, Ran and Amir for countless discussions from which he has benefited.  
C.C. would like to thank the PMIF group at ULB for its warm hospitality during the completion of this work. 
The work of I.S. is supported by the ERC STG grant number 335182, by the Israel Science Foundation under grant number 884/11. I.S. would also like to thank the United States-Israel Binational Science Foundation (BSF) for support under grant number 2010/629. In addition, the research of I.S. is supported by the I-CORE Program of the Planning and Budgeting Committee and by the Israel Science Foundation under grant number 1937/12. Any opinions, findings, and conclusions or recommendations expressed in this material are those of the authors and do not necessarily reflect the views of the funding agencies. 

\appendix

\section{Monopole Harmonics on the Sphere}\label{App: monopole harmonics}
Consider $S^2$ with angular coordinates $\theta \in [0, \pi]$, $\varphi \in [0, 2 \pi)$.
We introduce  the complex coordinate
\be
z= \tan{\theta\over 2} e^{i\varphi}~,
\ee
in the northern patch, which covers the sphere except for the south pole at $\theta= \pi$. The southern patch is similarly covered by a coordinate $z'$, with $z'= {1\over z}$ on the overlap $\theta \in (0, \pi)$. In this paper we always work on the northern patch, unless otherwise stated.

Consider the sphere with the round metric and one unit of magnetic flux~:
\be\label{metric S2}
ds^2 =  {4\over (1+ |z|^2)^2} dz d\bz~, \qquad A =  -{i\over 2(1 +|z|^2)}(\b z dz - z d\b z)~.
\ee
Here $A$ is the gauge field for the monopole.
This background is invariant under $SO(3)$ rotations. The metric in (\ref{metric S2}) had three Killing vectors,
\be\label{3 KV}
K_+ = z^2 \d_z + \d_\bz~, \qquad K_- = -\d_z -\bz^2 \d_\bz~, \qquad K_3= z\d_z -\bz\d_\bz~,
\ee
while the monopole gauge field is invariant along (\ref{3 KV}) up to gauge transformations,
\be\label{LieA mono}
\CL_{K_+} A  + {i\over 2} d z =0~, \qquad \CL_{K_-} A + {i\over 2} d \bz=0~, \qquad\CL_{K_3} A = 0~.
\ee
Here $\CL_K$ is the Lie derivative along $K$. Consequently, the $SO(3)$ transformations of  fields coupling to $A$ must be accompanied by gauge transformations, which are determined from (\ref{LieA mono}) up to integration constants. These integrations constants are fixed by the $SO(3)$ algebra
\be
[J_+, J_-]= 2 J_3~,\qquad [J_3, J_\pm]= \pm J_\pm~.
\ee
Therefore, the infinitesimal $SO(3)$ transformations on any field coupling to the monopole with electric charge $\rs$ are generated by
\be\label{L on fields}
J_- = \CL_{K_+} -{\rs\over 2} z~, \qquad J_+ = \CL_{K_-} -{\rs\over 2} \bz~, \qquad  J_3= \CL_{K_3} -{\rs\over 2}~.
\ee
Let us also define $J^2 = \half(J_+ J_- + J_- J_+)+ J_3^2$.

\subsection{Scalar Monopole Harmonics}
The scalar Laplacian on the monopole background (\ref{metric S2}) is given by
\be\label{S2 Lap scalar}
\Delta^\rs_{S^2} = -(1+ z \bar z)^2 \partial_z \partial_{\bar z} - \frac{\rs}{2}(1+ z \bar z) \left(z \partial_z - \bar z \partial_{\bar z} - \frac{\rs}{2} \right) - \frac{\rs^2}{4}~.
\ee
It acts on scalar fields of electric charge $\rs$. Due to the relation $\Delta^\rs_{S^2}+{\rs^2\over 4} = J^2 $,  we can diagonalize (\ref{S2 Lap scalar}) together with $J^2$, $J_3$~: 
\be
\Delta^\rs_{S^2}Y_{\rs\,jm} = \left( j(j+1)-\frac{\rs^2}{4} \right) Y_{\rs\,jm}~, \; J^2 Y_{\rs\,jm} = j(j+1)Y_{\rs\, jm}~,\; J_3 Y_{\rs\, jm} = m Y_{\rs\, jm}~.
\ee
Not all possible $SO(3)$ representations appear. The allowed values of $j$, $m$ are
\be\label{allowed eigenvalues scalar}
j= \frac{|\rs|}{2}, \frac{|\rs|}{2}+1,\ldots~, \quad\qquad m=-j,\ldots,j~.
\ee
Note that for $\rs$ odd the allowed values for the angular momentum are half-integer, therefore a scalar can behave like a fermion in a monopole background \cite{Wilczek:1981du}. 

The eigenfunctions $Y_{\rs\,jm}$ are known as monopole harmonics \cite{Wu:1976ge}. We will not need their explicit form, just the fact that they are orthonormal for fixed $\rs$,
\be
\int_{S^2}  d^2 x\sqrt{g} \, Y^\dagger_{rjm} Y_{rj'm'} = \delta_{jj'} \delta_{mm'}~.
\ee
Let us emphasize that the monopole harmonics are really sections of a non-trivial line bundle, with transitions functions 
\be
Y^{(N)}_{\rs\, jm} =\left( {z\over \bz}\right)^{{\rs\over 2}} Y^{(S)}_{\rs\,jm}
\ee
between the northern $(N)$ and southern $(S)$ patches.

\subsection{Spinor Monopole Harmonics}
Consider spinor fields on the sphere with a monopole. (See \cite{Abrikosov:2002jr,Benna:2009xd} for related discussions.) We choose the complex veilbein and gamma matrices on $S^2$~:
\be
 e = {2\over 1+ |z|^2} dz~,\quad \b e = {2\over 1+ |z|^2} d\bz~, \quad \gamma^{(e)} =  \mat{0 & 2 \cr  0 & 0}~, \quad \gamma^{(\b e)} =  \mat{0 & 0 \cr  2 & 0}~.
\ee
In this frame, the Dirac operator on  (\ref{metric S2}),  acting on a spinor of electric charge $\rs-1$, is given by
\be\label{Dirac op on S2}
-i \slashed{\nabla}^\rs_{S^2} = -i \begin{pmatrix} 0 & (1+z \bar z) \partial_z - \frac{1}{2}\rs \bar{z} \\  (1+ z \bar z) \partial_{\bar{z}}+ \frac{1}{2}(\rs-2)z & 0 \end{pmatrix}.
\ee
One can find its eigenvalues,
\be \label{eigenvalues Dirac}
-i\slashed{\nabla}^\rs_{S^2} \psi^{\pm}_{\rs-1\,jm} = \pm \lambda_{\rs j} \, \psi^{\pm}_{\rs-1\,jm}~,
\qquad \lambda_{\rs j} = \sqrt{\left( j+\frac{1}{2}\right)^2 - \frac{(\rs-1)^2}{4}}~,
\ee
where $j$, $m$ are the eigenvalues of $J^2$, $J_3$, respectively.
The allowed values are 
\be\label{allowed eigenvalues spinors}
j=\frac{|\rs-1|}{2}-\frac{1}{2}, \frac{|\rs-1|}{2}+\frac{1}{2},\ldots~, \qquad\quad m=-j,\ldots,j~,
\ee
with the lowest level for $j$ appearing only when $\rs \neq 1$. The lowest level $j=\frac{|\rs-1|}{2}-\frac{1}{2}$ corresponds to zero modes.
The eigenspinors $\psi^{\pm}_{\rs-1\,jm} $ can be expressed in terms of the scalar monopole harmonics. We have
\begin{align} \label{eigenfunctions}
(\psi^{\pm}_{\rs-1\, jm})_{\alpha} &= \frac{1}{\sqrt{2}} \begin{pmatrix}i Y_{\rs-2\, jm}  \\ \pm  Y_{\rs\, jm} \end{pmatrix}~,
\end{align} 
for the non-zero modes, $j > \frac{|\rs-1|}{2}-\frac{1}{2}$, and 
\begin{align}  \label{fer-0-modes}
(\psi_{\rs-1\, jm})_{\alpha} &= \begin{pmatrix}  Y_{\rs-2\,jm} \\0 \end{pmatrix}, & (\psi_{\rs-1\, jm})_{\alpha} &= \begin{pmatrix} 0\\ Y_{\rs\, jm}    \end{pmatrix}, 
\end{align}
for the zero modes, where on the left we have the  case $\rs > 1$ (with $j={\rs\over 2}-1$), and on the right the case $\rs < 1$ (with $j=-{\rs \over 2}$).
The eigenspinors are orthonormal for fixed $\rs$, 
\be
\int_{S^2}  d^2 x\sqrt{g} \,  (\psi^{\epsilon}_{\rs-1\, jm})^{\dagger} \psi^{\epsilon'}_{\rs-1\, j' m'} = \delta^{\epsilon \epsilon'} \delta_{j j'} \delta_{m m'},
\ee
where $\epsilon, \epsilon'= \pm 1$.

\section{Comments on the  Supersymmetric Pairing of Eigenvalues}\label{App: Pairing}
In this Appendix, we show that the operators \eqref{adj op LY LYt} are adjoint.  We also give an alternative argument for the supersymmetric pairing of eigenvalues discussed in section \ref{subsec: eigenvalues}.

\subsection{Adjoint Operators}
Let us prove that the operators $i\CLYt$ and $i\Omega^2 \CLY$ in \eqref{adj op LY LYt} are mutually adjoint. We have 
\be
Y^\mu A_\mu = -\frac{is}{2\Omega^2 c} \partial_{\b z} \log(\Omega^2 c s)~, \qquad\quad \b Y^\mu A_\mu = \frac{i}{2 s c} \partial_{z} \log \left(\frac{\Omega^4 c}{s} \right)~,
\ee
in the notation of section \ref{sec: susy with two supercharges}. It follows directly from the definition \eqref{def LKXY} that
\bea\label{LB Lphi formulas}
i\CLY B &= i s^{\frac{r}{2}}(\Omega^2 c)^{\frac{r}{2}-2}(\partial_{\b z} - \b h \partial_{\b w}) \left( \frac{B}{(\Omega^2c\,s)^{\frac{r}{2}-1}} \right)~, \cr
i\CLYt \phi &= i s^{\frac{r}{2}-1}\frac{\Omega^4}{(\Omega^4 c)^{\frac{r}{2}+1}}(\partial_{z} - h \partial_{w}) \left( \left(\frac{\Omega^4 c}{s}\right)^{\frac{r}{2}} \phi \right)~,
\eea
on fields $B$ and $\phi$ of $R$-charge $r-2$ and $r$, respectively.  
Recalling that $\Omega, c$ and $s$ are functions of $z, \bz$ only, with $\Omega$ and $c$ real, and that $\sqrt{g}={1\over 4}\Omega^4 c^2$, one can use \eqref{LB Lphi formulas} to prove \eqref{adjoint property} by direct computation.

\subsection{Another Derivation of the Eigenvalue Pairing}
Let us give another, complementary explanation of the claims of section \ref{subsec: eigenvalues} by exhibiting explicitly the supersymmetric pairing between bosonic and fermionic eigenmodes. Consider the eigenvalue equations
\be \label{eigenvalue_eq 2}
\Delta_{\rm bos} \phi = \Lambda_{\rm b} \phi~, \qquad\qquad 
 \Delta_{\rm fer} \Psi = \Lambda_{\rm f} \Psi~,
\ee
with the kinetic operators   \eqref{kinops new}. We want to compute the quantity
\be\label{def Zloc2}
Z_{\cM_4}^\Phi = {\det{\Delta_{\rm fer}} \over\det{\Delta_{\rm bos}}} = {\prod \Lambda_f \over \prod \Lambda_b}~.
\ee
Let us consider $\phi$ a bosonic solution  of \eqref{eigenvalue_eq 2},  which we  can also take to be an eigenstate of $i\CLK$ since $i\CLK$ commutes with $\Delta_{\rm bos}$. Let us moreover {\it assume} that the operator $i\Omega^2 \CLKt$ also commutes with $ \Delta_{\rm bos}$, so that we can diagonalize $\Delta_{\rm bos}, i\CLK$ and $i\Omega^2\CLKt$ simultaneously.~\footnote{We easily see that $[\Delta_{\rm bos}, \Omega^2 \CLKt]=0$ if and only if $F_{z w}= F_{\bz w}=0$, in addition to \eqref{constrain field strength FA}. We have $F_{z w}= \d_z A_w$, $F_{\bz w}= \d_\bz A_w$, and $A_{w}= \frac{3}{4}\kappa \Omega^2 - \frac{i}{2c^2} \partial_z \b h$. Recall that $\kappa$ is a function satisfying $K^{\mu}\partial_{\mu} \kappa =0$ but otherwise arbitrary. 
The partition function of a chiral multiplet is independent of $\kappa$ \cite{Closset:2013vra} and we are free to make a convenient choice. Even though such a choice is possible locally, there could be subtleties with setting $A_w$ to a constant globally.} 
We thus have
\begin{align}  \label{}
\Delta_{\rm bos} \phi &= \Lambda_{\rm b} \phi, & i \CLK \phi &= l_K \phi, & i \Omega^2 \CLKt \phi &= l_{\b K}  \phi \, .
\end{align}
We can use $\phi$ to construct its matching fermionic eigenstates. Define
\be  \label{def Psi paired}
\Psi_1  =  \mat{i \CLYt \phi \cr 0}~, \qquad\qquad 
\Psi_2 = \mat{0 \cr \phi }~.
\ee
 This mapping is  precisely the  one we would naively expect from the supersymmetry transformations \eqref{susychiralBC}. The fermions $\Psi_{1}, \Psi_2$ generate an invariant subspace of $ \Delta_{\rm fer}$, which reads
\begin{align}  \label{Dfer mat}
 \bmat{ \Delta_{\rm fer}}_{ \{ \Psi_1, \Psi_2 \}} = \bmat{ l_{K} & 1 \cr l_K l_{\b K} - \Lambda_{\rm b} & l_{\b K} } \, .
\end{align}
One can diagonalize this matrix to find two eigenvalues $\Lambda_{\rm f,\pm}$ whose product equals $\Lambda_b$.
Therefore, a bosonic eigenstate $\phi$ with eigenvalue $\Lambda_{\rm b}$ is generally paired with two fermionic eigenstates with eigenvalues $\Lambda_{\rm f,\pm}$, such that $\Lambda_{\rm f,+}\Lambda_{\rm f,-} = \Lambda_{\rm b}$.  One can easily invert this map. Therefore, the only contributions to \eqref{def Zloc2} come from unpaired modes, which are of two types.

Firstly, we can have a bosonic mode $\phi$ which is not completely cancelled by its fermionic partners. This happens if and only if  $\CLYt \phi=0$, and one can show that the net contribution to the partition function is the eigenvalue of the operator $i \CLK$~:
\be\label{eigenvaluesPhi 2}
i \CLK \phi = \lambda_\phi\, \phi\, , \qquad \quad \CLYt \phi =0~.
\ee
The eigenvalue $\lambda_\phi$ effectively contributes to the denominator of \eqref{def Zloc2}.
Secondly, we can have a  fermionic eigenmode with no paired boson, which must be of the type $\Psi_0= (B, 0)^T$.
Such a mode is a solution of the eigenvalue equations~:
\be\label{eigenvaluesB 2}
i \CLK B = \lambda_B\,  B\, , \qquad \quad \CLY B =0~.
\ee
The eigenvalue $\lambda_B$ contributes to the numerator of (\ref{def Zloc}). This proves \eqref{Zloc eigen}, \eqref{eigenvaluesBPhi}.

While this argument is very concrete, it is however less general than the discussion above \eqref{Z loc result} since we had to assume that  $[\Delta_{\rm bos}, \Omega^2 \CLKt]=0$.

\section{$S^3\times S^1$ Supersymmetric Backgrounds}\label{App: S3S1}
In order to preserve at least two supercharges on $S^3\times S^1$, we need to consider the following quotient of $\C^2-(0,0)$ \cite{Closset:2013vra}~:
\be\label{Hopf surface ii}
(z_1, z_2)\sim (p\, z_1, q\, z_2)~, \qquad 0< |p|\leq |q|<1~.
\ee
This quotient is a  complex manifold $\cM_4^{p,q}$ diffeomorphic to $S^3\times S^1$. Let us introduce the complex parameters $\sigma=\sigma_1 + i \sigma_2$ and $\tau=\tau_1 + i \tau_2$, defined by $p= e^{2\pi i \sigma}, q=e^{2\pi i \tau}$. It is also convenient to introduce the real coordinates $x, \theta, \varphi, \chi$, with $x\in [0, 2\pi )$ the $S^1$ coordinate and $\theta\in [0,\pi]$, $\varphi, \chi \in [0, 2\pi)$ the $S^3$ coordinates, with
\be
 z_1 = e^{i \sigma x} \cos{\theta\over 2}\, e^{i \varphi}~,\quad \qquad z_2 = e^{i\tau x} \sin{\theta\over 2} \,e^{i\chi}~.
\ee
The identification \eqref{Hopf surface ii} corresponds to $x\sim x+ 2\pi$. 
We consider the following Hermitian metric on $\cM_4^{p,q}$~:
\be\label{general met Hopf surf}
ds^2 = e^{2 \sigma_2 x} dz_1 d\bz_1 + e^{2 \tau_2 x} dz_2 d\bz_2~.
\ee
The background fields $V_\mu, A_\mu$ can be obtained from \eqref{2Q backgd ii} with $\sqrt{g}={1\over 4}e^{2(\sigma_2 +\tau_2)x}$, in the $z_1, z_2$ coordinates. In the complex frame $(e^1, e^2)=  (e^{ \sigma_2 x} dz_1, e^{\tau_2 x} d z_2)$, the Killing spinors are given by
\be\label{Killingspinors z1z2}
\zeta_\alpha = \sqrt{\frac{s}{2}} \left(  \begin{matrix}1 \cr   0 \end{matrix}\right)\, , \quad\qquad  
\t \zeta^\alphadot =   {1 \over \sqrt{2s}} {1\over \sqrt{\sigma_2 \tau_2}}  \left( \begin{matrix} i \sigma_2\, e^{2\sigma_2 x} \bz_1 \cr   -i \tau_2\, e^{\tau_2 x} \bz_2 \end{matrix}\right)~,
\ee
while the anti-homolorphic Killing vector \eqref{Kvec} is given by $K= -{i\over \sqrt{\sigma_2 \tau_2}}\left(\sigma_2 \bz_1 \d_{\bz_1} + \tau_2 \bz_2 \d_{\bz_2}\right)$. Note that $s$ must transform as
\be\label{transfo s as K sec S3S1}
s \sim e^{-2\pi i (\sigma_1 +\tau_1)} s
\ee
under the identification \eqref{Hopf surface ii}, so that the holomorphic two-form \eqref{P holo} is well-defined. We compensate \eqref{transfo s as K sec S3S1} by an $R$-symmetry transformation, and $s$ is then a scalar in the sense of \cite{Dumitrescu:2012ha}, which we can set to $1$. Consequently, any  field $\Phi$ of $R$-charge $r$ satisfies the twisted boundary condition
\be
\Phi \sim e^{i \pi r (\sigma_1 +\tau_1)}\Phi
\ee
as $x\sim x+ 2\pi$. Note that this condition depends on the choice of holomorphic coordinates.~\footnote{Upon a change of holomorphic coordinates, $s$ shifts by a phase (it transforms as $p g^{-{1\over 4}}$, with $p$ a section of the canonical line bundle $\cK$). That phase is removed by a $U(1)_R$ gauge transformation, which affects the boundary conditions of $R$-charged fields.} We will discuss a different choice of coordinates in the following, with different resulting boundary conditions for the fields.

In section \ref{subsec: S3 S1 Z}, we found convenient to use coordinates $w, z$ adapted to two supercharges in the sense of section \ref{sec: susy with two supercharges}, such that $K= \d_\bw$. A simple choice is
\be\label{def wz}
w= - i \sqrt{\tau_2 \over \sigma_2} \log z_1~, \qquad z={(z_2)^{\sqrt{\sigma_2\over \tau_2}}\over (z_1)^{\sqrt{\tau_2\over \sigma_2}}}
\ee
for the ``northern'' patch with $z_1\neq 0$, and $w'= w - i \log z, z'={1\over z}$ to cover the ``southern'' patch with $z_2 \neq 0$. These coordinates satisfy various identifications that follow from their definition and from \eqref{Hopf surface ii}. In the following we will focus on the case $\tau_2= \sigma_2$ for simplicity. In the general case, it is actually more convenient to use the $z_1, z_2$ coordinates and the metric \eqref{general met Hopf surf}.~\footnote{One can of course apply the formalism of section \ref{sec: chiral mult Z} in any coordinate system, solving the eingenvalue equations   \eqref{eigenvaluesBPhi}. Doing that in the $z_1, z_2$ coordinates, one recovers \eqref{Z S3S1 inf prod}. We will not present the details here, as it does not give anything new with respect to the $|p|=|q|$ subcase.}

\subsection{Background fields in $w,z$ coordinates, for $|p|=|q|$}
In the special case $\tau_2=\sigma_2$, the $w,z$ coordinates \eqref{def wz} simplify to \eqref{coord wz i}. The metric then takes the canonical form \eqref{2Q backgd i} with
\be
\Omega = 1~, \qquad h=  - {i\bz\over  1+|z|^2}~, \qquad \b h =  {iz\over  1+|z|^2}~,\qquad
c= {1\over 1+|z|^2}~.
\ee
This gives the Hermitian metric \eqref{metric S3S1 wz} considered in the main text.
The background fields $V_\mu, A_\mu$ take the simple form
\bea\label{A V S3S1 wz}
V_\mu dx^\mu  &=  \half (1 +  \kappa )(dw+ h dz) +\half( d\bw +\b h d\bz)~, \cr
A_\mu dx^\mu  &= \half h dz+ \half \b h d\bz +\frac{1}{2} \left(1 +{3\over 2} \kappa \right) ( dw + h dz)~.
\eea
The two Killing spinors take the form \eqref{Killingspinors}. In the special case $\tau_1 = -\sigma_1$, we can have four Killing spinors upon setting $\kappa= -2$ \cite{Dumitrescu:2012ha,Festuccia:2011ws}.~\footnote{For $\sigma_1+\tau_1\neq 0$ and $\kappa=-2$, we still have four Killing spinors locally but two of them are not globally defined.}  The present background has a $U(1)^3$ isometry in general, corresponding to the  three Killing vectors  $\d_x, \d_\varphi, \d_\chi$.
In these $w,z$ coordinates, $R$-charged fields satisfy the twisted boundary conditions
\be
\Phi \sim e^{\pi i r(\tau_1-\sigma_1)}\Phi~, \; (x\sim x+ 2\pi)~, \qquad \quad 
\Phi \sim e^{-2 \pi i r}\Phi~, \; (\varphi\sim \varphi+ 2\pi)~, 
\ee
as we circle the $x$ and $\varphi$ coordinates, respectively. 
(Note that the $S^1$ spanned by $\varphi$ does not shrink on the $w, z$ patch. A similar boundary condition  for $\chi\sim \chi +2\pi$ holds on the $w', z'$ patch.)
Correspondingly, the momentum operators along the $U(1)^3$ isometries are given by
\be
P_x = - i\d_x -{r\over 2}(\tau_1 -\sigma_1)~,\qquad
P_\varphi = - i\d_\varphi + r~, \qquad
P_\chi = - i\d_\chi~, 
\ee
for $R$-charged fields. This gives \eqref{P ops on S3S1} in the $w, z$ coordinates.

We also consider a background gauge field for a $U(1)_f$ symmetry as discussed in section \ref{subsec: backgd amu}. On $S^3\times S^1$, we can have background gauge fields corresponding to a one-dimensional family of holomorphic line bundles \cite{Closset:2013vra},
\be
a_\mu dx^\mu = -\half (a_r + i a_i) \omega^{1,0} -\half (a_r - i a_i) \omega^{0,1}~.
\ee 
Here $\omega^{1,0}$ is an element of $H^{1,0}(\cM_4^{p,q})$, which can be taken as $\omega^{1,0}=  \d(-2x)$. This gives \eqref{bckgd gauge field S3S1} in the main text. The parameter $a_r$ is a flat connection on the $S^1$, while the combination $a_r-i a_i$ is the holomorphic line bundle modulus entering the supersymmetric partition function \cite{Closset:2013vra}.

\bibliographystyle{utphys}
\bibliography{bibS2T2}{}

\providecommand{\href}[2]{#2}\begingroup\raggedright\begin{thebibliography}{10}

\bibitem{Dumitrescu:2012ha}
T.~T. Dumitrescu, G.~Festuccia, and N.~Seiberg, ``{Exploring Curved
  Superspace},'' \href{http://dx.doi.org/10.1007/JHEP08(2012)141}{{\em JHEP}
  {\bfseries 1208} (2012) 141},
\href{http://arxiv.org/abs/1205.1115}{{\ttfamily arXiv:1205.1115 [hep-th]}}.

\bibitem{Klare:2012gn}
C.~Klare, A.~Tomasiello, and A.~Zaffaroni, ``{Supersymmetry on Curved Spaces
  and Holography},'' \href{http://dx.doi.org/10.1007/JHEP08(2012)061}{{\em
  JHEP} {\bfseries 1208} (2012) 061},
\href{http://arxiv.org/abs/1205.1062}{{\ttfamily arXiv:1205.1062 [hep-th]}}.

\bibitem{Closset:2013vra}
C.~Closset, T.~T. Dumitrescu, G.~Festuccia, and Z.~Komargodski, ``{The Geometry
  of Supersymmetric Partition Functions},''
\href{http://arxiv.org/abs/1309.5876}{{\ttfamily arXiv:1309.5876 [hep-th]}}.

\bibitem{Suwa:1969}
T.~Suwa, ``{On ruled surfaces of genus 1},''
  \href{http://dx.doi.org/10.2969/jmsj/02120291}{{\em J. Math. Soc. Japan}
  {\bfseries 21} (1969) 291--311}.

\bibitem{Samtleben:2012gy}
H.~Samtleben and D.~Tsimpis, ``{Rigid supersymmetric theories in 4d Riemannian
  space},'' \href{http://dx.doi.org/10.1007/JHEP05(2012)132}{{\em JHEP}
  {\bfseries 1205} (2012) 132},
\href{http://arxiv.org/abs/1203.3420}{{\ttfamily arXiv:1203.3420 [hep-th]}}.

\bibitem{Festuccia:2011ws}
G.~Festuccia and N.~Seiberg, ``{Rigid Supersymmetric Theories in Curved
  Superspace},'' \href{http://dx.doi.org/10.1007/JHEP06(2011)114}{{\em JHEP}
  {\bfseries 1106} (2011) 114},
\href{http://arxiv.org/abs/1105.0689}{{\ttfamily arXiv:1105.0689 [hep-th]}}.

\bibitem{Imamura:2011su}
Y.~Imamura and S.~Yokoyama, ``{Index for three dimensional superconformal field
  theories with general R-charge assignments},''
  \href{http://dx.doi.org/10.1007/JHEP04(2011)007}{{\em JHEP} {\bfseries 1104}
  (2011) 007},
\href{http://arxiv.org/abs/1101.0557}{{\ttfamily arXiv:1101.0557 [hep-th]}}.

\bibitem{Hristov:2013spa}
K.~Hristov, A.~Tomasiello, and A.~Zaffaroni, ``{Supersymmetry on
  Three-dimensional Lorentzian Curved Spaces and Black Hole Holography},''
  \href{http://dx.doi.org/10.1007/JHEP05(2013)057}{{\em JHEP} {\bfseries 1305}
  (2013) 057},
\href{http://arxiv.org/abs/1302.5228}{{\ttfamily arXiv:1302.5228 [hep-th]}}.

\bibitem{Witten:1982df}
E.~Witten, ``{Constraints on Supersymmetry Breaking},''
\href{http://dx.doi.org/10.1016/0550-3213(82)90071-2}{{\em Nucl.Phys.}
  {\bfseries B202} (1982) 253}.

\bibitem{Witten:1986bf}
E.~Witten, ``{Elliptic Genera and Quantum Field Theory},''
\href{http://dx.doi.org/10.1007/BF01208956}{{\em Commun.Math.Phys.} {\bfseries
  109} (1987) 525}.

\bibitem{Benini:2013nda}
F.~Benini, R.~Eager, K.~Hori, and Y.~Tachikawa, ``{Elliptic genera of
  two-dimensional N=2 gauge theories with rank-one gauge groups},''
\href{http://arxiv.org/abs/1305.0533}{{\ttfamily arXiv:1305.0533 [hep-th]}}.

\bibitem{Benini:2013xpa}
F.~Benini, R.~Eager, K.~Hori, and Y.~Tachikawa, ``{Elliptic genera of 2d N=2
  gauge theories},''
\href{http://arxiv.org/abs/1308.4896}{{\ttfamily arXiv:1308.4896 [hep-th]}}.

\bibitem{Pestun:2007rz}
V.~Pestun, ``{Localization of gauge theory on a four-sphere and supersymmetric
  Wilson loops},'' \href{http://dx.doi.org/10.1007/s00220-012-1485-0}{{\em
  Commun.Math.Phys.} {\bfseries 313} (2012) 71--129},
\href{http://arxiv.org/abs/0712.2824}{{\ttfamily arXiv:0712.2824 [hep-th]}}.

\bibitem{Kapustin:2009kz}
A.~Kapustin, B.~Willett, and I.~Yaakov, ``{Exact Results for Wilson Loops in
  Superconformal Chern-Simons Theories with Matter},''
  \href{http://dx.doi.org/10.1007/JHEP03(2010)089}{{\em JHEP} {\bfseries 1003}
  (2010) 089},
\href{http://arxiv.org/abs/0909.4559}{{\ttfamily arXiv:0909.4559 [hep-th]}}.

\bibitem{Hama:2011ea}
N.~Hama, K.~Hosomichi, and S.~Lee, ``{SUSY Gauge Theories on Squashed
  Three-Spheres},'' \href{http://dx.doi.org/10.1007/JHEP05(2011)014}{{\em JHEP}
  {\bfseries 1105} (2011) 014},
\href{http://arxiv.org/abs/1102.4716}{{\ttfamily arXiv:1102.4716 [hep-th]}}.

\bibitem{Romelsberger:2007ec}
C.~Romelsberger, ``{Calculating the Superconformal Index and Seiberg
  Duality},''
\href{http://arxiv.org/abs/0707.3702}{{\ttfamily arXiv:0707.3702 [hep-th]}}.

\bibitem{Dolan:2008qi}
F.~Dolan and H.~Osborn, ``{Applications of the Superconformal Index for
  Protected Operators and q-Hypergeometric Identities to N=1 Dual Theories},''
  \href{http://dx.doi.org/10.1016/j.nuclphysb.2009.01.028}{{\em Nucl.Phys.}
  {\bfseries B818} (2009) 137--178},
\href{http://arxiv.org/abs/0801.4947}{{\ttfamily arXiv:0801.4947 [hep-th]}}.

\bibitem{Gerchkovitz:2013zra}
E.~Gerchkovitz, ``{Constraints on the R-charges of Free Bound States from the
  R\"omelsberger Index},''
\href{http://arxiv.org/abs/1311.0487}{{\ttfamily arXiv:1311.0487 [hep-th]}}.

\bibitem{Sohnius:1981tp}
M.~F. Sohnius and P.~C. West, ``{An Alternative Minimal Off-Shell Version of
  N=1 Supergravity},''
\href{http://dx.doi.org/10.1016/0370-2693(81)90778-4}{{\em Phys.Lett.}
  {\bfseries B105} (1981) 353}.

\bibitem{Komargodski:2010rb}
Z.~Komargodski and N.~Seiberg, ``{Comments on Supercurrent Multiplets,
  Supersymmetric Field Theories and Supergravity},''
  \href{http://dx.doi.org/10.1007/JHEP07(2010)017}{{\em JHEP} {\bfseries 1007}
  (2010) 017},
\href{http://arxiv.org/abs/1002.2228}{{\ttfamily arXiv:1002.2228 [hep-th]}}.

\bibitem{Dumitrescu:2011iu}
T.~T. Dumitrescu and N.~Seiberg, ``{Supercurrents and Brane Currents in Diverse
  Dimensions},'' \href{http://dx.doi.org/10.1007/JHEP07(2011)095}{{\em JHEP}
  {\bfseries 1107} (2011) 095},
\href{http://arxiv.org/abs/1106.0031}{{\ttfamily arXiv:1106.0031 [hep-th]}}.

\bibitem{Closset:2012ru}
C.~Closset, T.~T. Dumitrescu, G.~Festuccia, and Z.~Komargodski,
  ``{Supersymmetric Field Theories on Three-Manifolds},''
  \href{http://dx.doi.org/10.1007/JHEP05(2013)017}{{\em JHEP} {\bfseries 1305}
  (2013) 017},
\href{http://arxiv.org/abs/1212.3388}{{\ttfamily arXiv:1212.3388 [hep-th]}}.

\bibitem{Cassani:2013dba}
D.~Cassani and D.~Martelli, ``{Supersymmetry on curved spaces and
  superconformal anomalies},''
  \href{http://dx.doi.org/10.1007/JHEP10(2013)025}{{\em JHEP} {\bfseries 1310}
  (2013) 025},
\href{http://arxiv.org/abs/1307.6567}{{\ttfamily arXiv:1307.6567 [hep-th]}}.

\bibitem{AlvarezGaume:1983ig}
L.~Alvarez-Gaume and E.~Witten, ``{Gravitational Anomalies},''
\href{http://dx.doi.org/10.1016/0550-3213(84)90066-X}{{\em Nucl.Phys.}
  {\bfseries B234} (1984) 269}.

\bibitem{Atiyah:1955}
M.~F. Atiyah, ``Complex fibre bundles and ruled surfaces,''
  \href{http://dx.doi.org/10.1112/plms/s3-5.4.407}{{\em Proc. London Math. Soc.
  (3)} {\bfseries 5} (1955) 407--434}.

\bibitem{Benini:2012ui}
F.~Benini and S.~Cremonesi, ``{Partition functions of N=(2,2) gauge theories on
  S2 and vortices},''
\href{http://arxiv.org/abs/1206.2356}{{\ttfamily arXiv:1206.2356 [hep-th]}}.

\bibitem{Doroud:2012xw}
N.~Doroud, J.~Gomis, B.~Le~Floch, and S.~Lee, ``{Exact Results in D=2
  Supersymmetric Gauge Theories},''
  \href{http://dx.doi.org/10.1007/JHEP05(2013)093}{{\em JHEP} {\bfseries 1305}
  (2013) 093},
\href{http://arxiv.org/abs/1206.2606}{{\ttfamily arXiv:1206.2606 [hep-th]}}.

\bibitem{Wu:1976ge}
T.~T. Wu and C.~N. Yang, ``{Dirac Monopole Without Strings: Monopole
  Harmonics},''
\href{http://dx.doi.org/10.1016/0550-3213(76)90143-7}{{\em Nucl.Phys.}
  {\bfseries B107} (1976) 365}.

\bibitem{Witten:1993yc}
E.~Witten, ``{Phases of N=2 theories in two-dimensions},''
  \href{http://dx.doi.org/10.1016/0550-3213(93)90033-L}{{\em Nucl.Phys.}
  {\bfseries B403} (1993) 159--222},
\href{http://arxiv.org/abs/hep-th/9301042}{{\ttfamily arXiv:hep-th/9301042
  [hep-th]}}.

\bibitem{Kinney:2005ej}
J.~Kinney, J.~M. Maldacena, S.~Minwalla, and S.~Raju, ``{An Index for 4
  dimensional super conformal theories},''
  \href{http://dx.doi.org/10.1007/s00220-007-0258-7}{{\em Commun.Math.Phys.}
  {\bfseries 275} (2007) 209--254},
\href{http://arxiv.org/abs/hep-th/0510251}{{\ttfamily arXiv:hep-th/0510251
  [hep-th]}}.

\bibitem{Aharony:2003sx}
O.~Aharony, J.~Marsano, S.~Minwalla, K.~Papadodimas, and M.~Van~Raamsdonk,
  ``{The Hagedorn - deconfinement phase transition in weakly coupled large N
  gauge theories},'' {\em Adv.Theor.Math.Phys.} {\bfseries 8} (2004) 603--696,
\href{http://arxiv.org/abs/hep-th/0310285}{{\ttfamily arXiv:hep-th/0310285
  [hep-th]}}.

\bibitem{Benvenuti:2006qr}
S.~Benvenuti, B.~Feng, A.~Hanany, and Y.-H. He, ``{Counting BPS Operators in
  Gauge Theories: Quivers, Syzygies and Plethystics},''
  \href{http://dx.doi.org/10.1088/1126-6708/2007/11/050}{{\em JHEP} {\bfseries
  0711} (2007) 050},
\href{http://arxiv.org/abs/hep-th/0608050}{{\ttfamily arXiv:hep-th/0608050
  [hep-th]}}.

\bibitem{Feng:2007ur}
B.~Feng, A.~Hanany, and Y.-H. He, ``{Counting gauge invariants: The Plethystic
  program},'' \href{http://dx.doi.org/10.1088/1126-6708/2007/03/090}{{\em JHEP}
  {\bfseries 0703} (2007) 090},
\href{http://arxiv.org/abs/hep-th/0701063}{{\ttfamily arXiv:hep-th/0701063
  [hep-th]}}.

\bibitem{Chandrasekharan}
K.~{Chandrasekharan}, ``{Elliptic functions.}.'' {Grundlehren der
  Mathematischen Wissenschaften, 281. Berlin etc.: Springer-Verlag. XI, 189 p.
  DM 138.00}, 1985.

\bibitem{Closset:2012vg}
C.~Closset, T.~T. Dumitrescu, G.~Festuccia, Z.~Komargodski, and N.~Seiberg,
  ``{Contact Terms, Unitarity, and F-Maximization in Three-Dimensional
  Superconformal Theories},''
  \href{http://dx.doi.org/10.1007/JHEP10(2012)053}{{\em JHEP} {\bfseries 1210}
  (2012) 053},
\href{http://arxiv.org/abs/1205.4142}{{\ttfamily arXiv:1205.4142 [hep-th]}}.

\bibitem{Closset:2012vp}
C.~Closset, T.~T. Dumitrescu, G.~Festuccia, Z.~Komargodski, and N.~Seiberg,
  ``{Comments on Chern-Simons Contact Terms in Three Dimensions},''
  \href{http://dx.doi.org/10.1007/JHEP09(2012)091}{{\em JHEP} {\bfseries 1209}
  (2012) 091},
\href{http://arxiv.org/abs/1206.5218}{{\ttfamily arXiv:1206.5218 [hep-th]}}.

\bibitem{Jafferis:2010un}
D.~L. Jafferis, ``{The Exact Superconformal R-Symmetry Extremizes Z},''
  \href{http://dx.doi.org/10.1007/JHEP05(2012)159}{{\em JHEP} {\bfseries 1205}
  (2012) 159},
\href{http://arxiv.org/abs/1012.3210}{{\ttfamily arXiv:1012.3210 [hep-th]}}.

\bibitem{Hama:2010av}
N.~Hama, K.~Hosomichi, and S.~Lee, ``{Notes on SUSY Gauge Theories on
  Three-Sphere},'' \href{http://dx.doi.org/10.1007/JHEP03(2011)127}{{\em JHEP}
  {\bfseries 1103} (2011) 127},
\href{http://arxiv.org/abs/1012.3512}{{\ttfamily arXiv:1012.3512 [hep-th]}}.

\bibitem{Imamura:2011wg}
Y.~Imamura and D.~Yokoyama, ``{N=2 supersymmetric theories on squashed
  three-sphere},'' \href{http://dx.doi.org/10.1103/PhysRevD.85.025015}{{\em
  Phys.Rev.} {\bfseries D85} (2012) 025015},
\href{http://arxiv.org/abs/1109.4734}{{\ttfamily arXiv:1109.4734 [hep-th]}}.

\bibitem{Alday:2013lba}
L.~F. Alday, D.~Martelli, P.~Richmond, and J.~Sparks, ``{Localization on
  Three-Manifolds},''
\href{http://arxiv.org/abs/1307.6848}{{\ttfamily arXiv:1307.6848 [hep-th]}}.

\bibitem{QuineZeta}
J.~R. Quine, S.~H. Heydari, and R.~Y. Song, ``{Zeta Regularized Products},''
  {\em Transactions of the American Mathematical Society} {\bfseries 338}
  no.~1, (1993) 213--231.

\bibitem{Romelsberger:2005eg}
C.~Romelsberger, ``{Counting chiral primaries in N = 1, d=4 superconformal
  field theories},''
  \href{http://dx.doi.org/10.1016/j.nuclphysb.2006.03.037}{{\em Nucl.Phys.}
  {\bfseries B747} (2006) 329--353},
\href{http://arxiv.org/abs/hep-th/0510060}{{\ttfamily arXiv:hep-th/0510060
  [hep-th]}}.

\bibitem{Ruijs}
S.~Ruijsenaars, ``{On Barnes' multiple zeta and gamma functions},''
  \href{http://dx.doi.org/10.1006/aima.2000.1946}{{\em Advances in Mathematics}
  {\bfseries 156 (1)} (2000) 107–132}.

\bibitem{FriedmanRuijSB}
E.~Friedman and S.~Ruijsenaars, ``{Shintani--Barnes zeta and gamma
  functions},'' \href{http://dx.doi.org/10.1016/j.aim.2003.07.020}{{\em
  Advances in Mathematics} {\bfseries 187 (2)} (2004) 362–395}.

\bibitem{Nawata:2011un}
S.~Nawata, ``{Localization of N=4 Superconformal Field Theory on S1 x S3 and
  Index},'' \href{http://dx.doi.org/10.1007/JHEP11(2011)144}{{\em JHEP}
  {\bfseries 1111} (2011) 144},
\href{http://arxiv.org/abs/1104.4470}{{\ttfamily arXiv:1104.4470 [hep-th]}}.

\bibitem{Wilczek:1981du}
F.~Wilczek, ``{Magnetic Flux, Angular Momentum, and Statistics},''
\href{http://dx.doi.org/10.1103/PhysRevLett.48.1144}{{\em Phys.Rev.Lett.}
  {\bfseries 48} (1982) 1144}.

\bibitem{Abrikosov:2002jr}
J.~Abrikosov, A.A., ``{Dirac operator on the Riemann sphere},''
\href{http://arxiv.org/abs/hep-th/0212134}{{\ttfamily arXiv:hep-th/0212134
  [hep-th]}}.

\bibitem{Benna:2009xd}
M.~K. Benna, I.~R. Klebanov, and T.~Klose, ``{Charges of Monopole Operators in
  Chern-Simons Yang-Mills Theory},''
  \href{http://dx.doi.org/10.1007/JHEP01(2010)110}{{\em JHEP} {\bfseries 1001}
  (2010) 110},
\href{http://arxiv.org/abs/0906.3008}{{\ttfamily arXiv:0906.3008 [hep-th]}}.

\end{thebibliography}\endgroup

\end{document}